\documentclass[aps,showpacs,preprintnumbers,amsmath,amssymb]{revtex4}
\UseRawInputEncoding
\usepackage{dcolumn}
\usepackage[dvips]{epsfig}
\usepackage{graphicx}
\usepackage{amssymb}
\usepackage{color}
\usepackage{enumerate}
\usepackage{epsfig}

\def \be {\begin{equation}}
\def \ee {\end{equation}}
\def \bea {\begin{eqnarray}}
\def \eea {\end{eqnarray}}

\begin{document}
\baselineskip=0.8 cm
\title{\bf Tidal effects of dark matter halo around a galactic black hole }
\author{Jiayi Liu$^{1}$, Songbai Chen$^{1,2}$\footnote{Corresponding author: csb3752@hunnu.edu.cn}, Jiliang
Jing$^{1,2}$ \footnote{jljing@hunnu.edu.cn}}

\affiliation{ $ ^1$ Department of Physics,  Synergetic Innovation Center for Quantum Effects and Applications, Hunan
Normal University,  Changsha, Hunan 410081, People's Republic of China
\\
$ ^2$Center for Gravitation and Cosmology, College of Physical Science and Technology, Yangzhou University, Yangzhou 225009, People's Republic of China}

\begin{abstract}
\baselineskip=0.5 cm
\begin{center}
{\bf Abstract}
\end{center}

We have investigated the tidal forces and geodesic deviation motion in the spacetime of a black hole in the galaxy with dark matter halo. Our results show that the tidal force and geodesic deviation
motion depend on the dark matter halo mass  and the typical lengthscale of galaxy. The effect of the typical lengthscale of galaxy on  tidal force is opposite to that of dark matter mass. For the radial tidal force,
with the increasing mass of dark matter,  it increases in the region far from the black hole, but decreases in the region near black hole. For the angular tidal force, its absolute value of angular tidal force monotonously increases with the dark matter halo mass. Especially, the angular tidal force also depends on the particle's energy and the effects of dark matter become more distinct for the test particle with high energy, which is different from those in the usual static black hole spacetimes. We also present the change of geodesic deviation vector with the dark matter halo mass  and the typical lengthscale of galaxy under two kinds of initial conditions.

\end{abstract}

 \pacs{ 04.70.-s, 04.70.Bw, 97.60.Lf }
\maketitle
\newpage

\section{Introduction}

Dark matter is a kind of mysterious invisible substance with only gravitational interaction, which may theoretically exist in the Universe based on cosmic observation data including the cosmic microwave background radiation and baryon acoustic oscillations \cite{Bullock:2017xww, Planck:2015fie}. It is believed that dark matter makes up around $85\%$ of all the matter in the Universe. However, the nature of dark matter is still unclear. Thus, it is one of the outstanding challenges for us to understand the properties of dark matter.

The first-ever image of black hole, released by the Event Horizon Telescope (EHT) Collaboration, confirms that there is a supermassive black hole at the center of  elliptical galaxy M87 \cite{EventHorizonTelescope:2019dse,EventHorizonTelescope:2019uob}, which opens a new era of testing gravity in strong field regime. Moreover, recent astronomical observations \cite{deBlok:2008wp,Guo:2015epa}, such as  the spiral galaxy rotation curve and mass-luminosity ratio of elliptical galaxy, indicate that dark matter may cluster at the center of galaxies and close to black holes. Thus, it is important and interesting to study the interaction between dark matter and black hole. If there is dark matter in the vicinity of the black hole, then the dark matter and its distribution could modify the geometry of black hole, which could lead to some new observable effects arising from dark matter. Kiselev \cite{Kiselev:2002dx} obtained a static black hole solution surrounded by a specific matter. As the equation of state is equal to zero, this solution describes the modification of black hole metric caused by dark matter. The Kiselev's black hole solution has also been generalized to the high dimension black hole case \cite{Chen:2008ra} and the rotating black hole case \cite{Ghosh:2015ovj}. Making use of a mass function related to some distribution matter, some black hole metrics disformed by dark matter have been obtained \cite{Xu:2018wow, Xu:2020jpv,Zhang:2022roh,Liu:2021xfb,Jusufi:2020cpn,Hou:2018bar,Konoplya:2019sns, Sadeghian_2013}. Recently, Cardoso \textit{et al} \cite{Cardoso:2021wlq} obtained an interesting static black hole solution by solving Einstein equations with the energy momentum tensor originating from the
galactic matter, which describes a galactic black hole immersed in the dark matter
halo with Hernquist distribution. With this solution, the effects of dark mater halo on quasinormal modes, scattering and optical phenomena  have been studied in \cite{Konoplya:2021ube,Zhang:2021bdr,Jusufi:2022jxu,Stuchlik:2021gwg,b2019,Zou_2020}. The galactic  black hole solution surrounded by dark matter halo with other distributions have been discussed in \cite{Konoplya:2022hbl}.

Tidal disruption is one of the most magnificent phenomena  in galaxies, which occurs as a star passes sufficiently close to a supermassive black hole so that the tidal forces can destroy the star. The electromagnetic emission caused in tidal disruption events can help us to probe the feature of the corresponding supermassive black hole. Thus, there are a lot of efforts to study tidal effects including shape deformation of a body in black hole spacetimes.
In a Schwarzschild black hole spacetime, it is found that the tidal force makes a body falling towards black hole stretched in the radial direction and compressed in the angular one \cite{Hong:2020bdb}.
In the Reissner-Nordstr\"{o}m case \cite{Crispino:2016pnv}, the charge of the black hole affects sharply the tidal force so that its radial and angular components change their signs as the body falls
towards to the center of black hole. The tidal effects have been
investigated in some other static spacetimes
including regular black holes \cite{Lima:2020wcb,Sharif:2018gaj}, Kiselev black holes
\cite{Shahzad:2017vwi}, and naked singularity \cite{Goel:2015zva}. Moreover, the effect of Gauss-Bonnet coupling constant on
tidal forces and geodesic deviation vector has been studied in four-dimensional Gauss-Bonnet black hole spacetime \cite{Li:2021izh}. The investigation also shows \cite{Wheeler:1971ler} that in ergosphere of a rotating black hole a star broken  by tidal interaction can emit a
jet composed of the debris, which could provide a potential mechanism to explain the formation of jets near black holes.
Subsequently, the study of tidal effects have also been
performed in various black holes \cite{Kesden:2011ee,1985MNRAS.212L,1973ApJF,2005PhRvD,2019ApJH,2020EPJPL,Cardoso:2020hca}. The main purpose
of this paper is to study the tidal effects in the galactic black hole with a dark matter halo \cite{Cardoso:2021wlq}
and to probe how dark matter mass and galaxy lengthscale  affect tidal forces and the
motion of geodesic deviation vector.

The paper is organized as follows: In Sect.II, we briefly
review the galactic black hole solution with a dark matter halo  \cite{Cardoso:2021wlq} and the
geodesics equation. In Sect. III, we investigate tidal forces on
a body falling free along radial direction in the galactic black hole solution with a dark matter halo. In Sect. IV, we present the dynamical evolution of deviation vector in the spacetime of a black hole with dark matter halo
and analyze the effects from dark matter halo and galaxy lengthscale. Finally, we end the
paper with a summary.

\section{Geodesics in a spacetime of a galactic black hole with a dark matter halo}

\begin{figure}
\includegraphics[width=8cm]{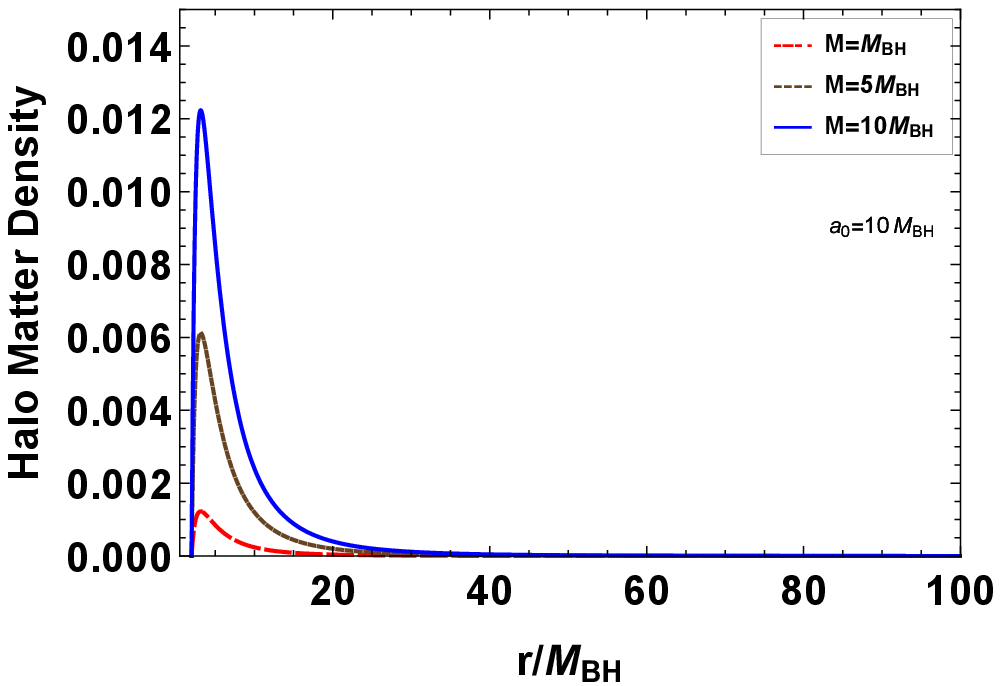}
\includegraphics[width=8cm]{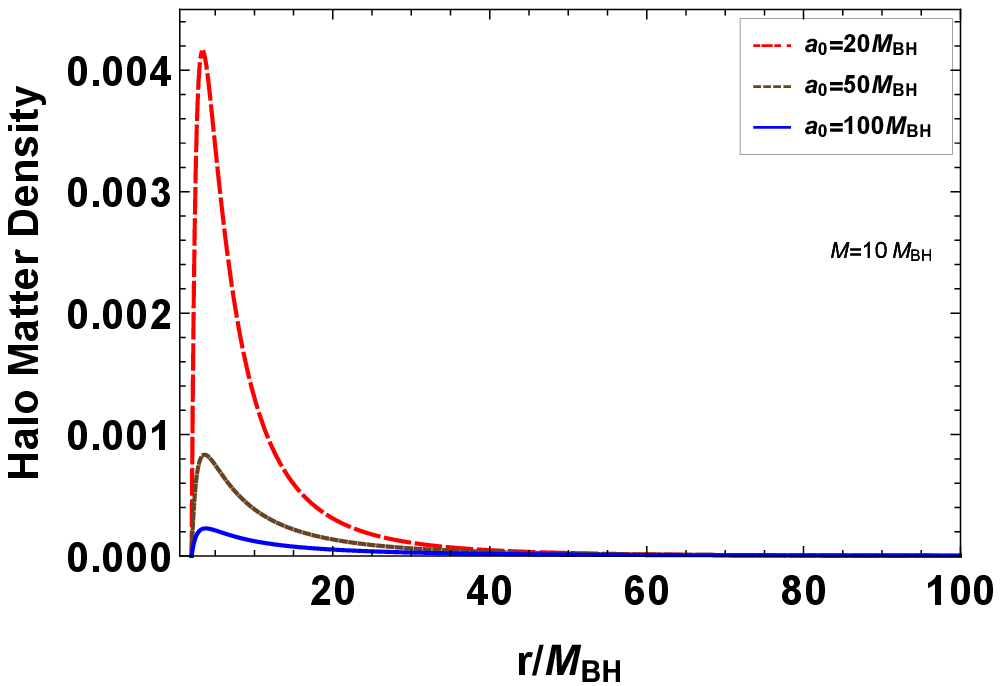}
\caption{  Change of the density distribution with  the halo mass $M$ and the lengthscale parameter $a_0$ for  the black hole spacetime (\ref{metric}). }\label{f1}
\end{figure}

Let us now to review briefly the galactic black hole solution with a dark matter halo obtained in \cite{Cardoso:2021wlq}. It is well known that the Hernquist-type density distribution \cite{Hernquist_1991} can be used to describe the S\'{e}rsic profiles observed in bulges and elliptical galaxies,
 \begin{equation}
  \rho=\frac{a_0 M}{2 \pi  r \left(a_0+r\right)^3},\label{matt1}
\end{equation}
where $M$ is the total mass of the dark matter halo and $a_0$ is a typical lengthscale of the galaxy. With such a density distribution of matter, an interesting exact solution \cite{Cardoso:2021wlq} was introduced to describe a black hole immersed in the center of a galaxy.
The metric of the black hole solution has a form
\begin{equation}
  \text{ds}^2= -f(r)\text{dt}^2 +\frac{1}{1-\frac{2 m(r)}{r}}\text{dr}^2 +r^2 \left(\text{d$\theta$ }^2+ \sin^2\theta\text{d$\phi$}^2\right),\label{metric}
\end{equation}
where the mass distribution of matter around black hole is  \cite{Cardoso:2021wlq}
\begin{equation}
   m(r)=M_{\text{BH}}+\frac{M r^2}{\left(a_0+r\right){}^2} \left(1-\frac{2 M_{\text{BH}}}{r}\right){}^2.
 \end{equation}
Here $M_{\text{BH}}$ is the mass of black hole in the center. The function $f(r)$ can be expressed as \cite{Cardoso:2021wlq}
\begin{equation}
f(r)=\left(1-\frac{2 M_{\text{BH}}}{r}\right) e^{\Upsilon (r)},
\end{equation}
with
\begin{eqnarray}
&&\Upsilon (r)=-\pi  \sqrt{\frac{M}{\xi }} +2 \sqrt{\frac{M}{\xi }} \arctan \left(\frac{a_0-M+r}{\sqrt{M \xi }}\right),\hfill\label{u7}\\
&&\xi=2 a_0+4 M_{\text{BH}}-M.\label{u8}
\end{eqnarray}
It must be pointed out that in the solution (\ref{metric}) the dark matter surrounding the black hole is assumed to own only non-zero tangential pressures and vanishing radial pressure. The spacetime (\ref{metric}) is asymptotically flat and the corresponding ADM mass is $M+M_{\text{BH}}$. It has a horizon at $r_{H}=2M_{\text{BH}}$ and a curvature singularity at $r=0$. Eqs.(\ref{u7}) and (\ref{u8}) means that $2 a_0+4 M_{\text{BH}}-M>0$, which is consistent with the best fit of galaxies' observations  where $a_0> 10^4M$ \cite{Navarro:1995iw}.
In the spacetime of a black hole immersed dark matter (\ref{metric}), the corresponding density function of dark matter becomes
\begin{equation}
\rho = \frac{2 M ( a_0 +2 M_{\text{BH}}) \left(1-\frac{2 M_{\text{BH}}}{r}\right)}{r ( a_0 +r)^3},\label{matt2}
\end{equation}
which differs from the original Hernquist-type density distribution (\ref{matt1}) due to the interaction between dark matter and the black hole at the center of galaxy.  As the mass of black hole $M_{\text{BH}}$ tends to zero, the density function (\ref{matt2}) reduces to the original one (\ref{matt1}). At the horizon $r_{H}=2M_{\text{BH}}$, the matter density vanishes.
In fact, the parameters $M$, $a_0$ and $M_{\text{BH}}$ determine the geometry of spacetime (\ref{metric}). Observation of galaxies corresponds to the regime $a_0>10^4M$ \cite{Navarro:1995iw}. Moreover, as $a_0\gg M_{\text{BH}}$, the density function of dark matter (\ref{matt2}) at large distances becomes the Hernquist-type density distribution (\ref{matt1}). Thus, as in \cite{Cardoso:2021wlq,Konoplya:2021ube}, we focus on the regime where $M_{\text{BH}}\ll M\ll a_0$. Especially, in this parameter region, the spacetime (\ref{metric}) has only a curvature singularity at $r=0$ and  the another singularity $r=M-a_0\pm\sqrt{M(M-2a_0-4M_{\text{BH}})}$ do not exist since $M-2a_0-4M_{\text{BH}}<0$, which means that the spacetime (\ref{metric}) is regular in the physical region $r>r_H$. Moreover, the choice of such a parameter region also leads to that the Ricci scalar near the horizon $R\sim M/(a^2_0M_{\text{BH}})$ can be made small in a controlled way, and decays as $R \sim4Ma_0/r^4$ at large distances \cite{Cardoso:2021wlq}. In Fig. (\ref{f1}), we also present the change of dark matter density distribution with the halo mass $M$ and the lengthscale parameter $a_0$. It is shown that the density increases with the mass $M$ for the fixed $a_0=10 M_{\text{BH}}$, but decreases with $a_0$ for the fixed $M=10 M_{\text{BH}}$.

In the black hole spacetime (\ref{metric}), the  geodesic motion equation of a test massive  particle along the radial direction is \cite{1984General}
\begin{equation}
f(r) \dot{t}^2-\frac{\dot{r}^2}{1-\frac{2 m(r)}{r}}=1,\label{raeq1}
\end{equation}
where the dot represents the differentiation with respect to the proper time $\tau$. It is obvious that a massive particle moving along the geodesics in the spacetime (\ref{metric}) owns two conserved
quantities, i.e., the energy $E$ and angular momentum $L$. As in \cite{Martel_2002}, with the equation $E=f(r) \dot{t}$, the equation (\ref{raeq1}) can be expressed as
\begin{equation}
\dot{r}^2=\left(\frac{E^2}{f(r)}-1\right) \left(1-\frac{2 m(r)}{r}\right).
\label{radal2}
\end{equation}
The Newtonian radial acceleration \cite{Symon1971Mechanics} for the particle in black hole surrounded by dark matter halo (\ref{metric})  can be expressed as
\begin{eqnarray}
A^R=\ddot{r}&=&\frac{rm'(r)-m(r)}{2r^2}-\frac{E^2}{2f(r)}\bigg[\frac{rm'(r)-m(r)}{r^2}
+\frac{f'(r)}{2f(r)}\bigg(1-\frac{2 m(r)}{r}\bigg)\bigg]\nonumber\\&=&
-\frac{M_{\text{BH}}}{r^2}-\frac{1}{(r+a_0)^3}\bigg\{\frac{(r-2M_{\text{BH}})M}{r^2}\bigg[r^2-(6M_{\text{BH}}+a_0)r
-2M_{\text{BH}}a_0\bigg]\nonumber\\
  &&+2E^2 M(2M_{\text{BH}} +a_0)e^{\sqrt{\frac{M}{2a_0+4M_{\text{BH}}-M}}[(\pi-2\arctan{\frac{r-M+ a_0}{\sqrt{M (2a_0+4M_{\text{BH}}-M)}}}]}\bigg\}.
\end{eqnarray}
Obviously, the absolute value of Newtonian radial acceleration increases with
the energy $E$ of the test particle, which differs from that in the Schwarzschild and Reissner-Nordstr\"{o}m  black hole spacetimes where the Newtonian radial acceleration is independent of test particle's energy $E$.
\begin{figure}
\includegraphics[width=6.0cm]{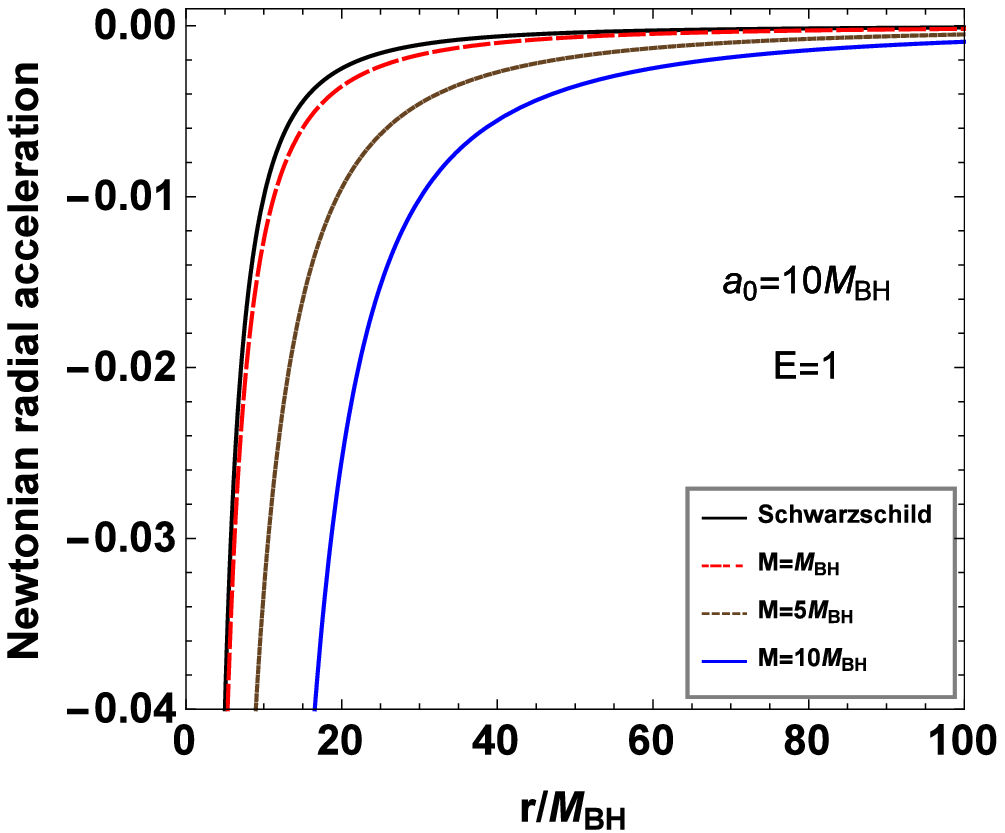}\includegraphics[width=6.0cm]{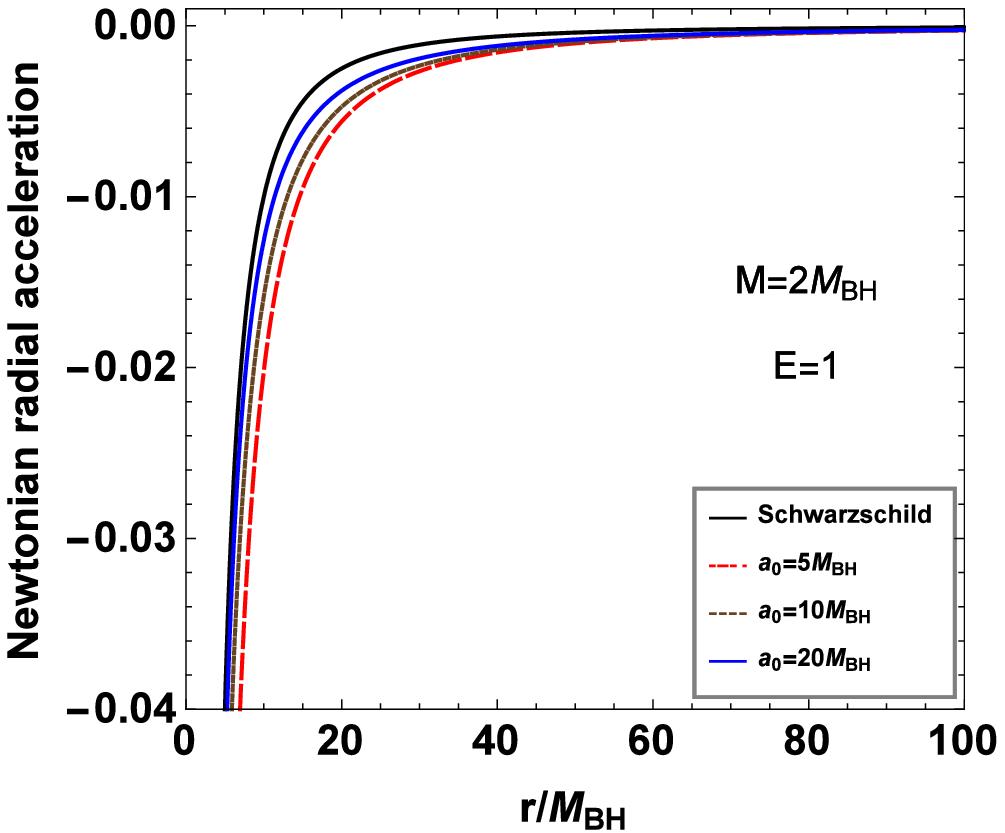}\\
\includegraphics[width=6.0cm]{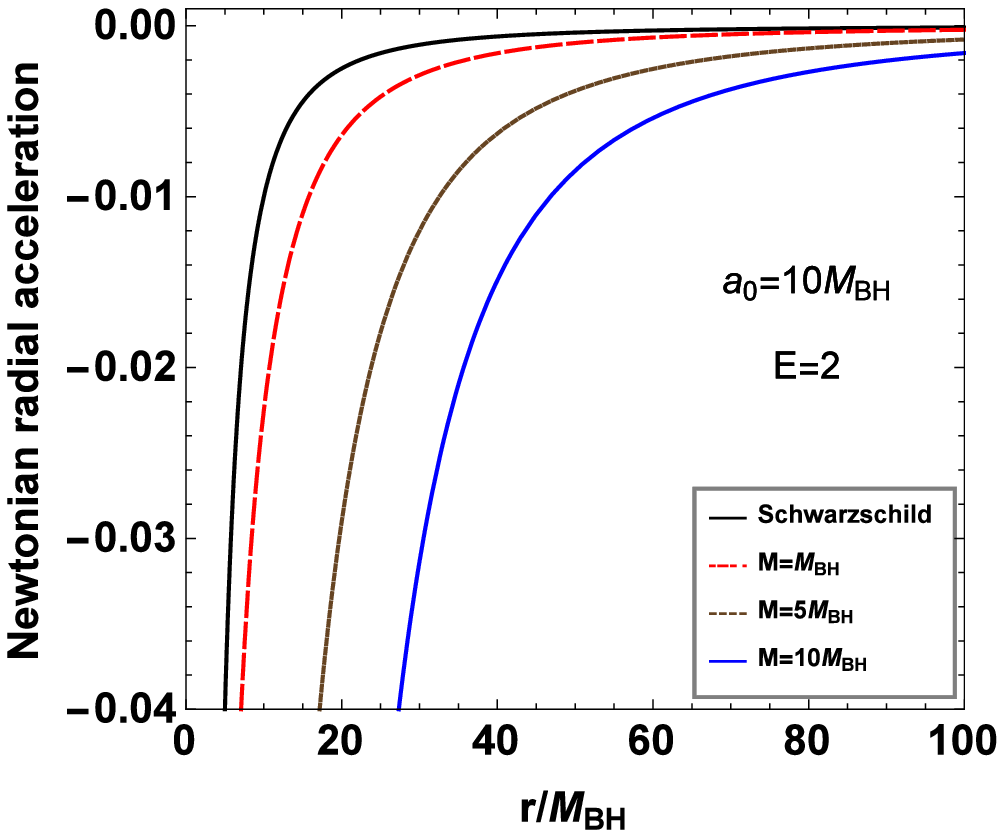}\includegraphics[width=6.0cm]{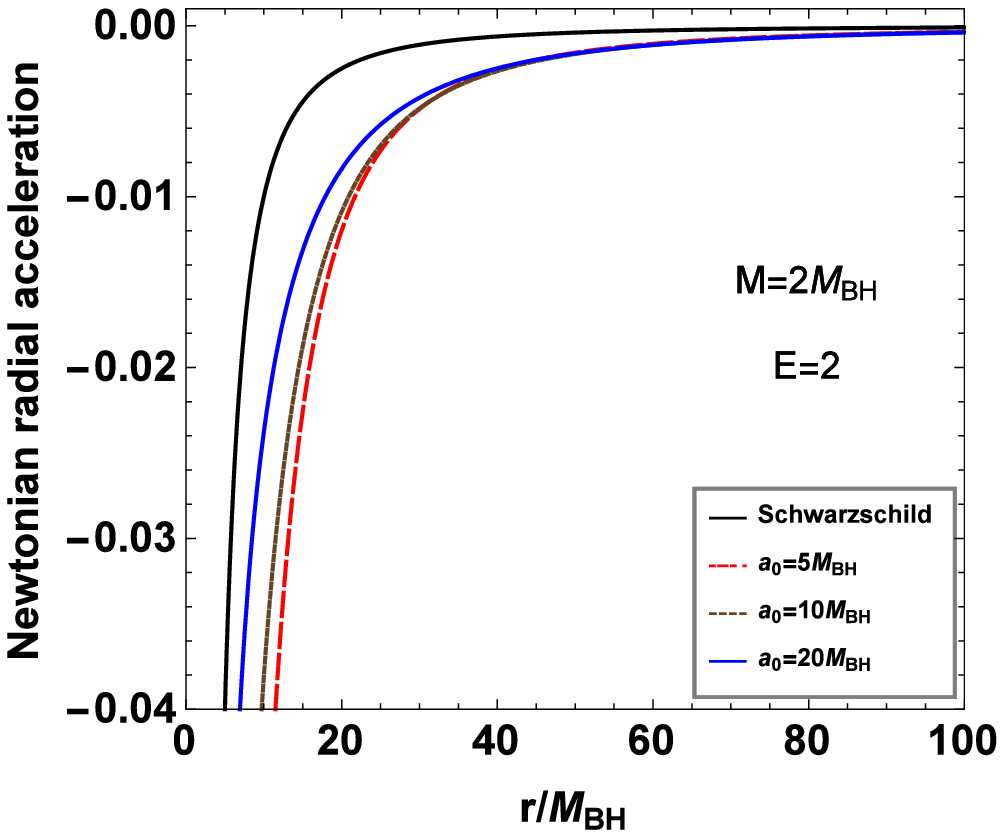}
\caption{ Change of Newtonian radial acceleration with parameters $M$ and $a_0$ for  a test particle moving along the radial direction in the background of a galactic black hole with dark matter halo (\ref{metric}).}
\label{f2s}
\end{figure}
 In Fig. (\ref{f2s}), we present the change of Newtonian radial acceleration with parameters $M$ and $a_0$ for  a test particle moving along the radial direction in the black holes with dark matter halo (\ref{metric}). It is shown that the absolute value of Newtonian radial acceleration increases with the dark matter mass $M$, but decreases with the lengthscale parameter $a_0$ of the galaxy. Moreover, we also find that effects of the parameters $M$ and $a_0$ on the Newtonian radial acceleration become more remarkable in the case of the test particle with higher energy $E$. From Eq. (\ref{radal2}), the we find that the test particle falling freely from rest at $r=b>r_H$ does not
bounce back in its radial motion since Eq. (\ref{radal2}) does not own another root satisfied $\dot{r}=0$ under the previous condition $M<2 a_0+4 M_{\text{BH}}$, which is similar to that in the Schwarzschild case.

\section{Tidal force of a neutralbody in radial free fall in the background of a galactic black hole with dark matter halo}

 Let us now to study the tidal forces on the massive particle in the background of a galactic black hole with dark matter halo (\ref{metric}). It is well known that the tidal forces are governed by a geodesic
deviation equation of a spacelike vector $\eta^{\mu}$ describing the
distance between two infinitesimally close particles following
geodesics \cite{1993Introducing,2007General}. The geodesic deviation equation is
\begin{equation}\label{cedpl}
  \frac{D^2 \eta ^{\mu }}{\text{$D$$\tau $}^2}-R_{\sigma \nu \rho}^{\mu } v^{\sigma} v^{\nu  }\eta ^{\rho }  =0,
\end{equation}
where $v^{\nu }$ is the unit vector tangent to the geodesic. In the spacetime of a black hole with dark matter halo (\ref{metric}), the tetrad basis related to a freely falling frame can be expressed as
\begin{equation}
\begin{aligned}
  \hat{e}_{\hat{0}}^{\mu }&=\bigg(\frac{E}{f(r)}, -\sqrt{\left(\frac{E^2}{f(r)}-1\right) \left(1-\frac{2 m(r)}{r}\right)}, 0, 0\bigg), \hfill\\
  \hat{e}_{\hat{1}}^{\mu }&=\bigg(-\frac{1}{f(r)}\sqrt{\frac{E^2}{f(r)}-1}, \frac{E}{f(r)} \sqrt{f(r) \left(1-\frac{2 m(r)}{r}\right)}, 0, 0\bigg), \hfill\\
  \hat{e}_{\hat{2}}^{\mu }&=(0, 0, \frac{1}{r}, 0),\hfill\\
  \hat{e}_{\hat{3}}^{\mu }&=(0, 0, 0, \frac{1}{r \sin \theta }),\hfill\\
  \end{aligned}\label{tetrad}
\end{equation}
which obeys the condition
\begin{equation}
  \hat{e}_{\alpha  \hat{\nu }} \hat{e}_{\hat{\mu }}^{\alpha }=\eta _{\hat{\nu } \hat{\mu }},
\end{equation}
Here $\eta _{\hat{\nu } \hat{\mu }}$ is the Minkowski metric. The indices with and without hat correspond to the tetrad and coordinate basis indices, respectively. Moreover, we have $\hat{e}_{\hat{0}}^{\mu }=v^{\mu }$. With the tetrad basis (\ref{tetrad}), the geodesic deviation vector $\eta^{\mu }$ can be expanded as $\eta^{\mu }=\hat{e}_{\hat{\nu}}^{\mu } \eta ^{\hat{\mu }}$  with the fixed  component $ \eta ^{\hat{0}}=0$.
In the background of a black hole with dark matter halo (\ref{metric}), the non-vanishing independent components of the Riemann tensor are \cite{1984General}
\begin{eqnarray}
R^1{}_0{}_1{}_0&=&\frac{1}{2}f''(r)\left(1-\frac{2 m(r)}{r}\right)+\frac{1}{4}\bigg[f'(r)\left(1-\frac{2 m(r)}{r}\right)'-\left(1-\frac{2 m(r)}{r}\right)\frac{f'(r)^2}{f(r)}\bigg],\hfill\nonumber \\
  R^1{}_2{}_1{}_2&=&\frac{m(r)-rm'(r)}{r},\quad \quad \quad    R^1{}_3{}_1{}_3=\frac{m(r)-rm'(r)}{r} \sin ^2\theta ,\hfill \nonumber\\
  R^2{}_3{}_2{}_3&=&\frac{2 m(r) \sin ^2\theta}{r},\quad \quad \quad  R^2{}_0{}_2{}_0=R^3{}_0{}_3{}_0=\frac{f'(r)}{2 r}\left(1-\frac{2 m(r)}{r}\right).
\end{eqnarray}
Making use of the relationship $R_{\hat{\beta} \hat{\gamma} \hat{\delta}}^{\hat{\alpha}}=e_{\mu}^{\hat{\alpha}} e_{\hat{\beta}}^{\nu} e_{\hat{\gamma}}^{\rho} e_{\hat{\delta}}^{\sigma} R_{\nu \rho \sigma}^{\mu}$ , one can get the Riemann curvature tensor in the tetrad basis
\begin{eqnarray}\label{ff-Riemann}
R_{\;\hat{1} \hat{0} \hat{1}}^{\hat{0}}&=&-\frac{1}{4 f'(r)}\bigg[\frac{f'(r)^2}{f(r)} \left(1-\frac{2 m(r)}{r}\right) \bigg]',\nonumber\\
R_{\;\hat{2} \hat{0} \hat{2}}^{\hat{0}}&=& R_{\;\hat{3} \hat{0}\hat{3}}^{\hat{0}}=\frac{1}{2r}\bigg[E^2\bigg(\frac{r-2 m(r)}{rf(r)}\bigg)'-\bigg(1-\frac{2 m(r)}{r}\bigg)'\bigg].
\end{eqnarray}
Here, we also note that the components $R_{\;\hat{2} \hat{0} \hat{2}}^{\hat{0}}$ and $R_{\;\hat{3} \hat{0}\hat{3}}^{\hat{0}}$ depend on the energy of particle, which could lead to the some new features of tidal force in the spacetime of a black hole with dark matter halo (\ref{metric}).
Substituting above formula into the geodesic deviation equation (\ref{cedpl}), one can find that the equations for tidal forces in radial free-fall reference frames satisfy \cite{Abdel_Megied_2005}.
\begin{eqnarray}
  \ddot{\eta }^{\hat{1}}&=&\frac{1}{r^3}\bigg\{2M_{\text{BH}}+\frac{(r-2M_{\text{BH}})M}{(r+a_0)^3}\bigg[2r^2+(a_0-6M_{\text{BH}})r-
  4M_{\text{BH}}a_0\bigg] \nonumber\\&&
  -\frac{Mr^2(2M_{\text{BH}}+a_0)}{(r+a_0)[r^2+2(a_0-M)r+4MM_{\text{BH}}+a^2_0]}\bigg\} \eta ^{\hat{1}},
\label{tidalrf}
\end{eqnarray}
and
\begin{eqnarray}
  \ddot{\eta}^{\hat{i}}&=&-
  \bigg\{\frac{M_{\text{BH}}}{r^3}+\frac{(r-2M_{\text{BH}})M}{r^3(r+a_0)^3}\bigg[r^2-(6M_{\text{BH}}+a_0)r-2M_{\text{BH}}a_0\bigg]\nonumber\\
  &&+\frac{2E^2 M(2M_{\text{BH}} +a_0)}{r(r+a_0)^3}e^{\sqrt{\frac{M}{2a_0+4M_{\text{BH}}-M}}[(\pi-2\arctan{\frac{r-M+ a_0}{\sqrt{M (2a_0+4M_{\text{BH}}-M)}}}}]\bigg\}\eta^{\hat{i}},
  \label{tidalanf}
\end{eqnarray}
where $i= 2,3$ correspond to the angular coordinate $\theta$ and $\phi$, respectively.
Obviously, the radial and angular tidal forces  depend on the black hole mass $M_{\text{BH}}$, the mass of the dark matter halo $M$ and the typical lengthscale $a_0$. Especially, the angular tidal force also is related to the energy $E$ of the test particle in the spacetime. As the parameter $M$ vanishes or $a_0$ tends infinite, the equations (\ref{tidalrf}) and (\ref{tidalanf}) reduce to those in Schwarzschild black hole spacetime, which can be explained by a fact that  the dark matter density (\ref{matt2}) vanishes in these two limit cases.
\begin{figure}
\includegraphics[width=6cm]{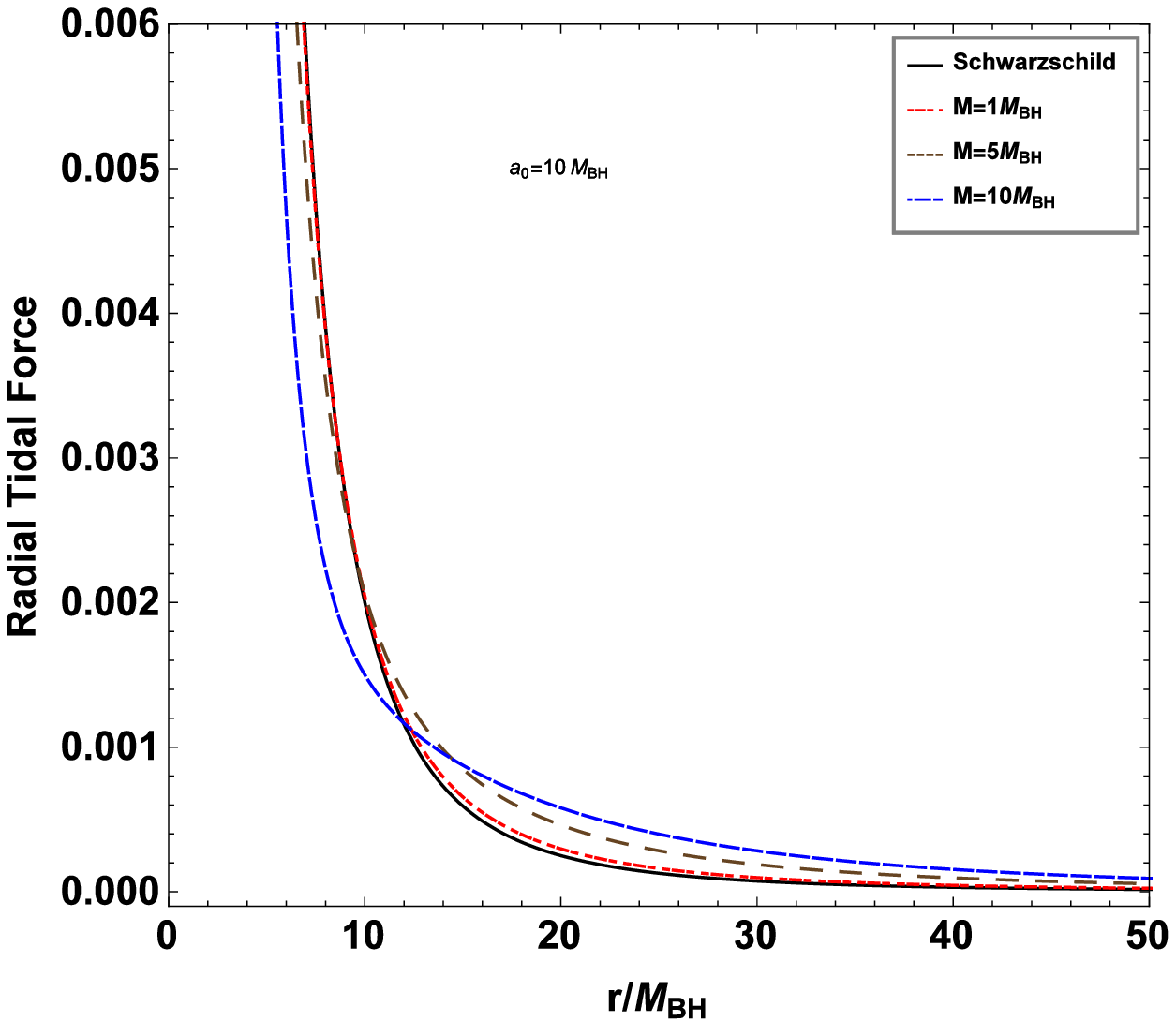}
\includegraphics[width=6cm]{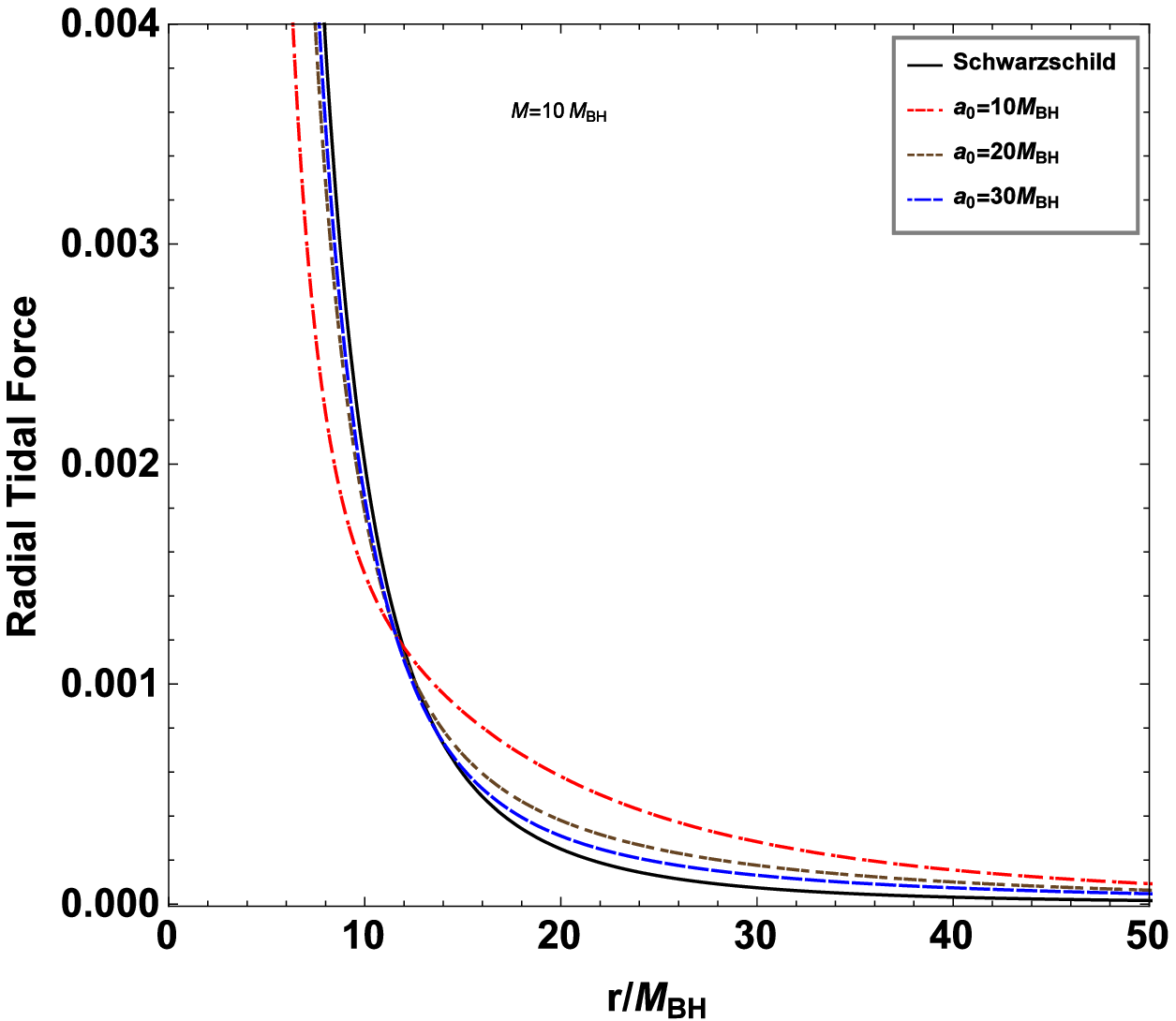}
\caption{Change of the radial tidal force with the dark matter mass parameter $M$ and  the typical lengthscale $a_0$  in the background of a galactic black hole with dark matter halo. }\label{rdtf1}
\end{figure}

\begin{figure}
\includegraphics[width=6cm]{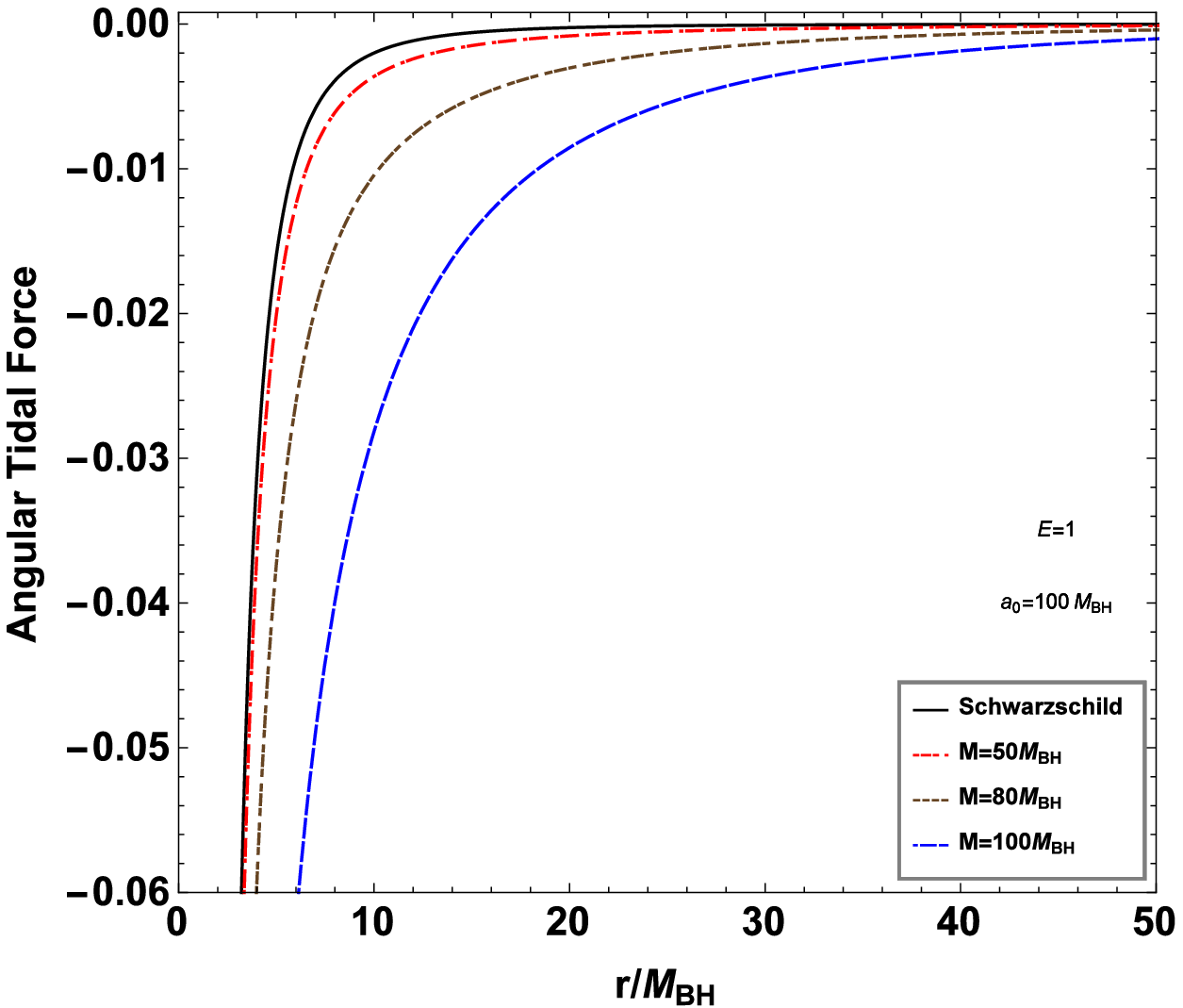}
\includegraphics[width=6cm]{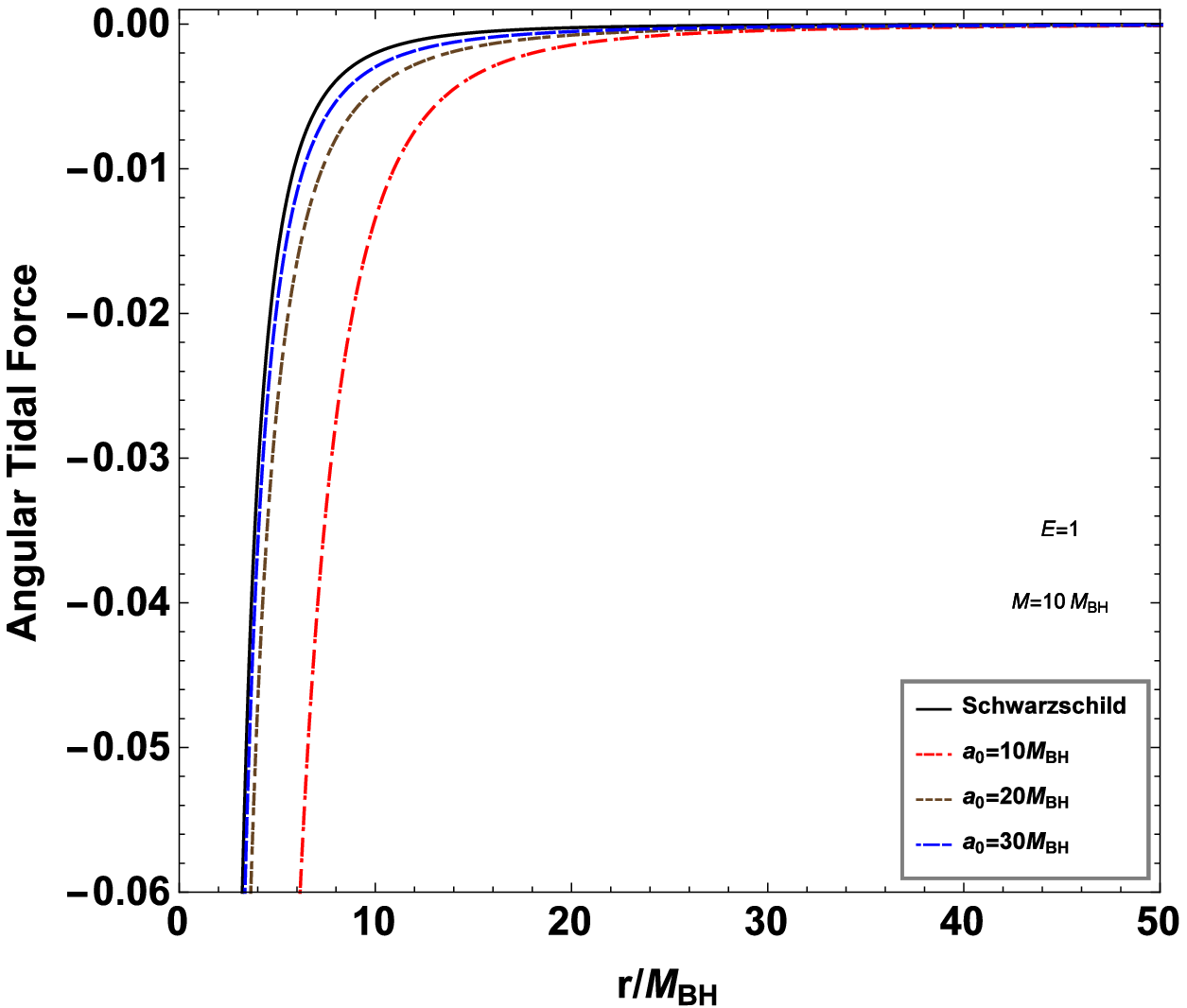}\\
\includegraphics[width=6cm]{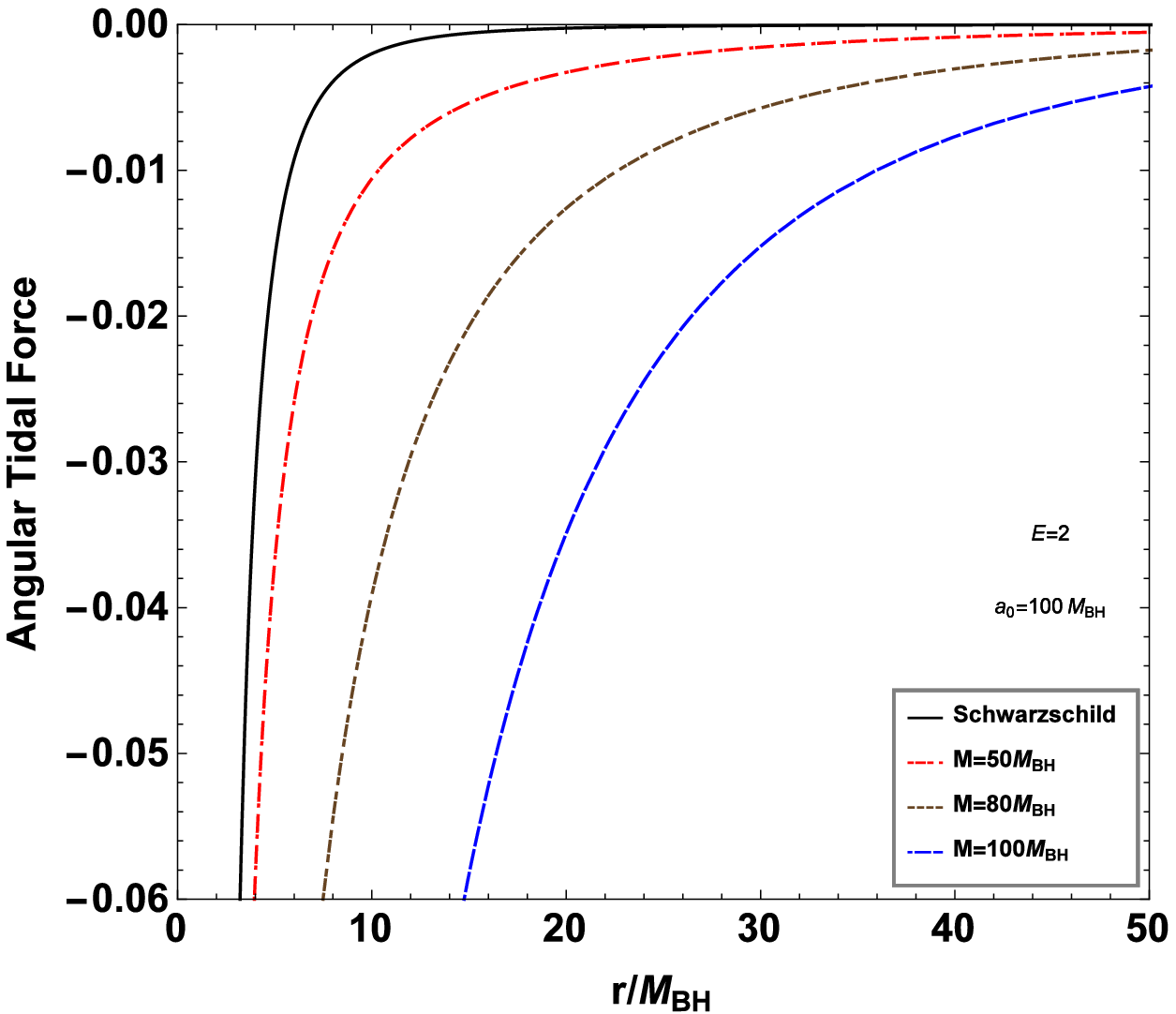}
\includegraphics[width=6cm]{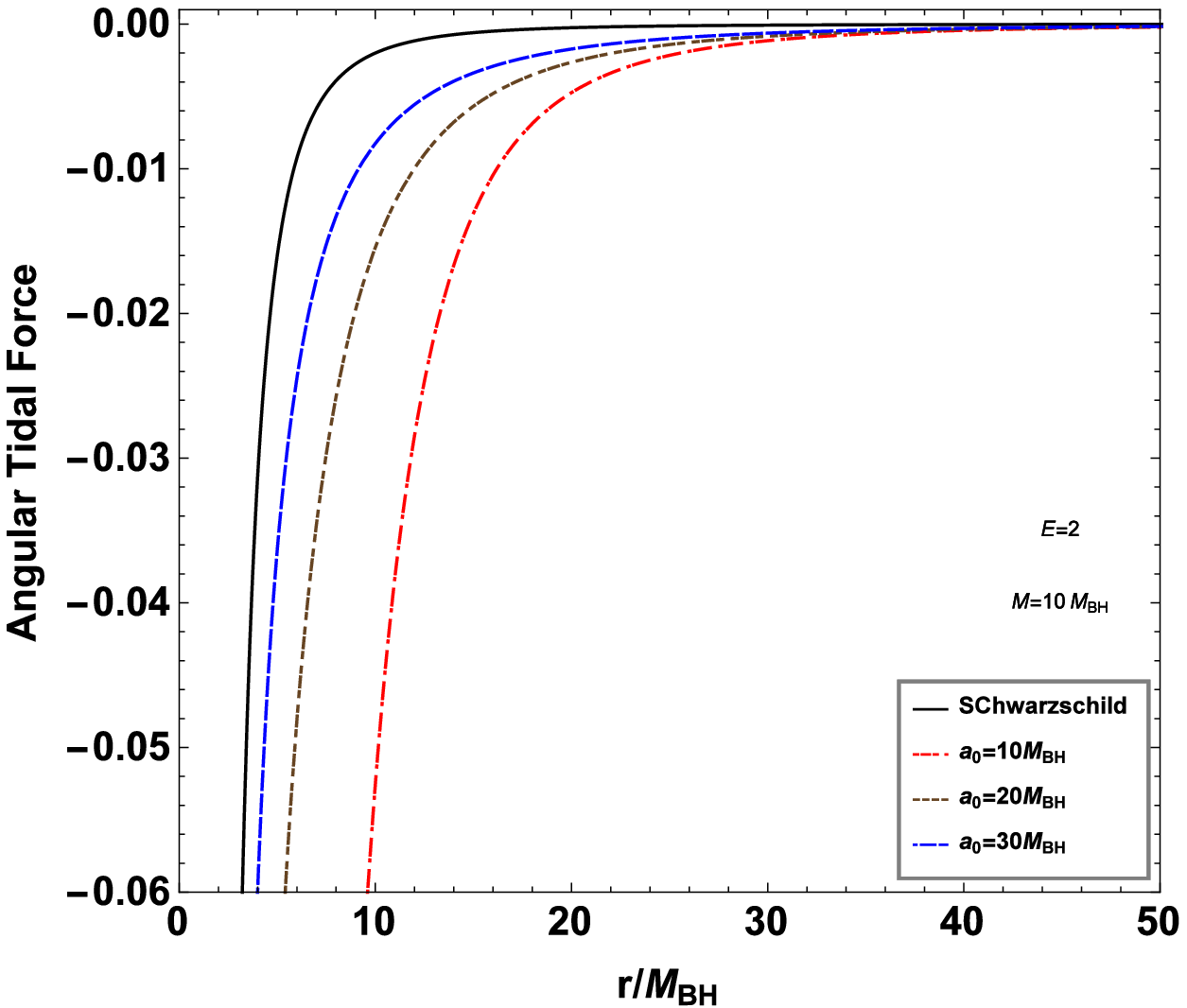}
\caption{Change of the angular tidal force with the dark matter mass parameter $M$ and  the typical lengthscale $a_0$  in the background of a galactic black hole with dark matter halo. }\label{angutf1}
\end{figure}

The changes of tidal forces with the dark matter mass parameter $M$ and  the typical lengthscale $a_0$ of galaxy are also shown in Figs. (\ref{rdtf1}) and (\ref{angutf1}).
For the fixed $a_0$, one can find that with the increasing $M$, the radial tidal force increases in the region far from the black hole, but decreases in the region near black hole. It is understandable since for the test particle in the far region the gravity arising from the dark matter around black hole has the same direction as that of black hole at the center of galaxy, which makes the gravitational field stronger than in the case without dark matter. But for the particle in the near region, they are in the opposite directions, which makes the gravitational field weaker than in the pure Schwarzschild case.
Moreover, we also find that the effect of $a_0$ on radial tidal force is opposite to that of $M$ in the background of a black hole with dark matter halo. The main reason is that the dark matter density (\ref{matt2}) increases with the mass parameter $M$, but decreases with the scale parameter $a_0$.
Fig. (\ref{angutf1}) shows that the angular tidal force is negative for all $M$ and $a_0$, which is consistent with that in the Schwarzschild black hole case.  The absolute value of angular tidal force monotonously increases with the dark matter halo mass, but decreases with the typical lengthscale $a_0$ of galaxy.  Especially, we also find that the angular tidal force also depends on the particle's energy $E$ and  the effects of $M$ and $a_0$ become more distinct for the test particle with high energy $E$, which is different from those in the usual static black hole spacetimes where the angular tidal force is independent of the test particle's energy $E$.

\section{Evolution of deviation vector in the spacetime of a black hole with dark matter halo}

In this section we solve the geodesic deviation equations (\ref{tidalrf}) and (\ref{tidalanf}) and analyse the dynamical evolution of the deviation vector $\eta ^{\hat{\mu}}$ for a particles radially free-falling in the spacetime of a black hole with dark matter halo (\ref{metric}).
Making use of the equation
\begin{equation}
dr/{\text{d$\tau $}}=-\sqrt{\left(1-\frac{2 m(r)}{r}\right)\left(\frac{E^2}{f(r)}-1\right)},
\end{equation}
it is easy to find that the geodesic deviation equations (\ref{tidalrf}) and (\ref{tidalanf}) can be rewritten as
\begin{eqnarray}
 &&\left(\frac{E^2}{f(r)}-1\right)\left(1-\frac{2 m(r)}{r}\right) \frac{d^2\eta^{\hat{1}}}{dr^2}
 +\frac{1}{2}\bigg[ E^2 \bigg(\frac{r-2 m(r)}{rf(r)}\bigg)' - \left(1-\frac{2 m(r)}{r}\right)'\bigg]\frac{d\eta^{\hat{1}}}{dr}\nonumber\\
 &&+\frac{1}{4 f'(r)}\bigg[\frac{f'(r)^2}{f(r)} \left(1-\frac{2 m(r)}{r}\right)\bigg]' \eta ^{\hat{1}}=0,\hfill \label{ddfene1}\\
&&\left(\frac{E^2}{f(r)}-1\right)\left(1-\frac{2 m(r)}{r}\right) \frac{d^2\eta^{\hat{i}}}{dr^2}
 +\frac{1}{2}\bigg[ E^2 \bigg(\frac{r-2 m(r)}{rf(r)}\bigg)'-\left(1-\frac{2 m(r)}{r}\right)'\bigg]\frac{d\eta^{\hat{i}}}{dr}\nonumber\\
&&-\frac{1}{2r}\bigg[ E^2 \bigg(\frac{r-2 m(r)}{rf(r)}\bigg)'-\left(1-\frac{2 m(r)}{r}\right)'\bigg]\eta^{\hat{i}}=0.
\end{eqnarray}
The general solution for the angular component $\eta^{\hat{i}}$ can be expressed as
\begin{eqnarray}
\eta^{\hat{i}}=r\bigg[C_1+C_2\int \frac{dr}{r^2\sqrt{\bigg(\frac{E^2}{f(r)}-1\bigg)\bigg(1-\frac{2 m(r)}{r}}\bigg)}\bigg],\label{so2}
\end{eqnarray}
where the coefficients $C_1$ and $C_2$ are the constants of integration. However, for the radial component $\eta^{\hat{1}}$, we can not find such kind of analytical solutions. Thus, we must solve numerically the differential equation (\ref{ddfene1}). As in \cite{Crispino:2016pnv}, we here consider two types of initial conditions describing dust of particles starting at the region outside the event horizon $r=b>r_{H}$.
The first type of initial conditions ICI is
\begin{equation}
 \eta ^{\hat{\alpha}}(b)=1,\quad\quad\quad\dot{\eta}^{\hat{\alpha}}(b)=0,                                            \end{equation}
which corresponds to a particle-like dust released from rest at $r=b$. This means that the four-velocity component of particle $\dot{r}=0$ and then the energy of the dust is fixed to $E=f(b)$. The second kind of initial condition ICII is
\begin{equation}
 \eta ^{\hat{\alpha}}(b)=0, \quad\quad\quad\dot{\eta}^{\hat{\alpha}}(b)=1,                                            \end{equation}
 which denotes to a  dust ``exploding" at the starting point $r=b$.
\begin{figure}
\includegraphics[width=4.0cm]{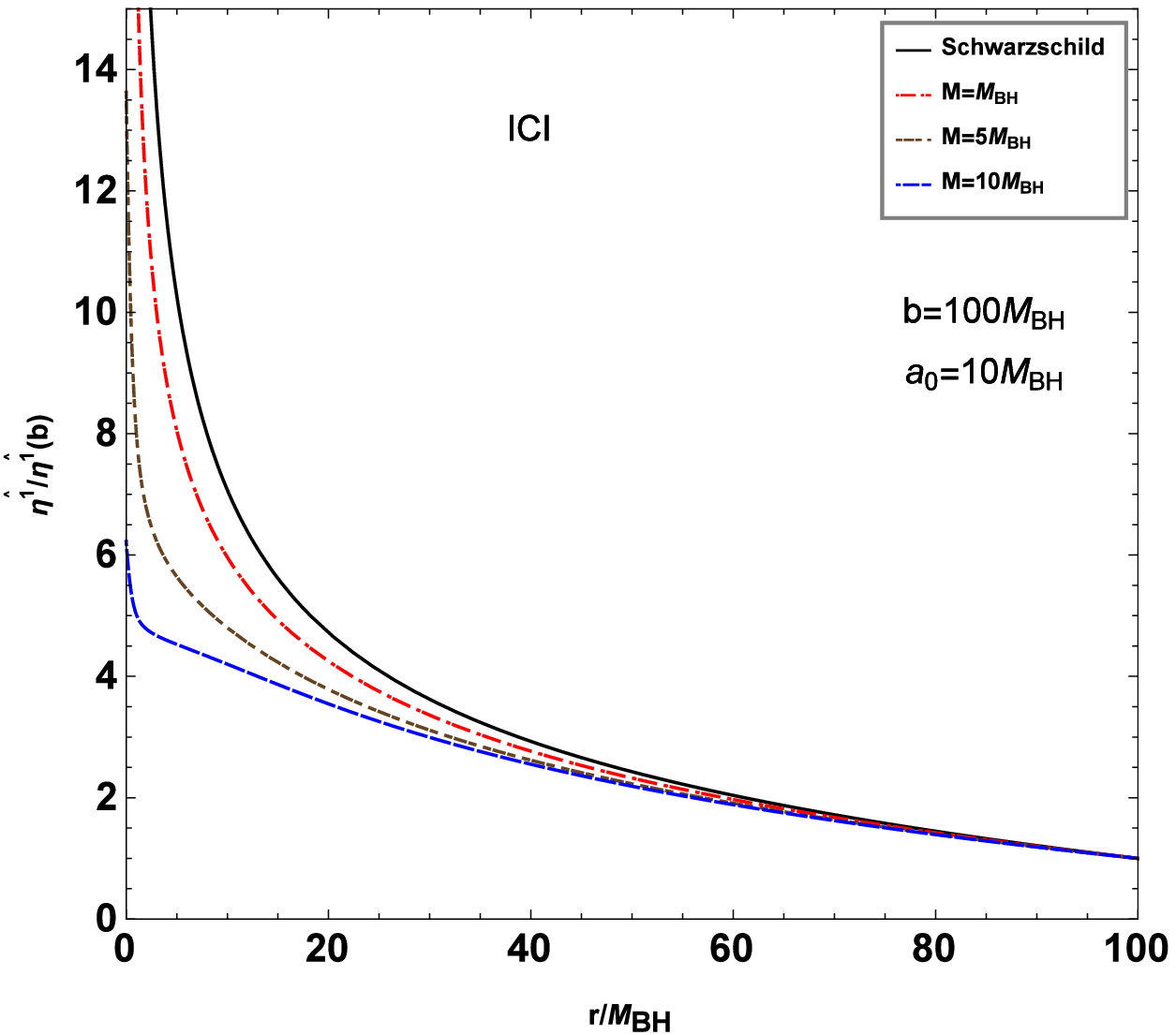}
\includegraphics[width=4.0cm]{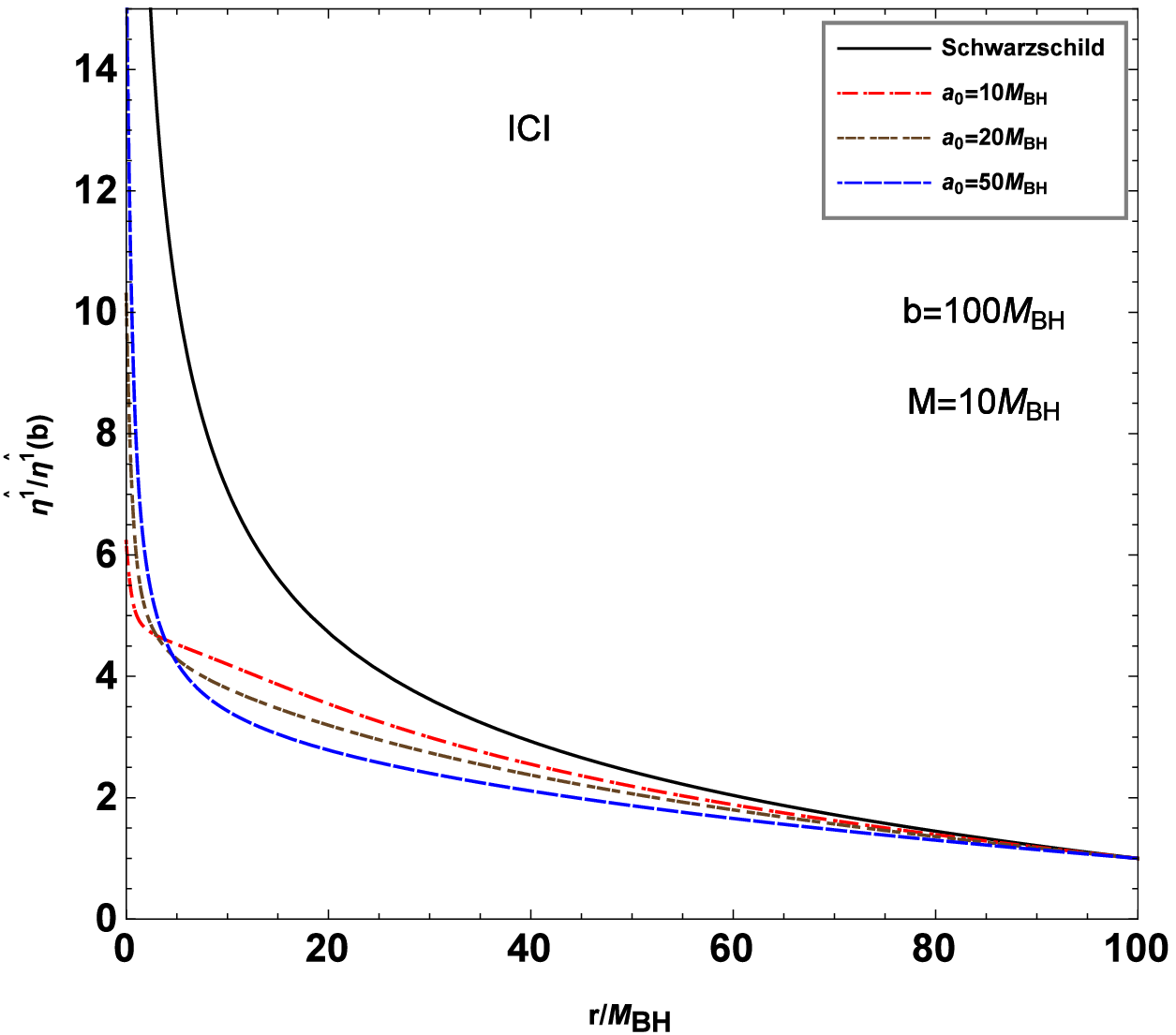}
\includegraphics[width=4.0cm]{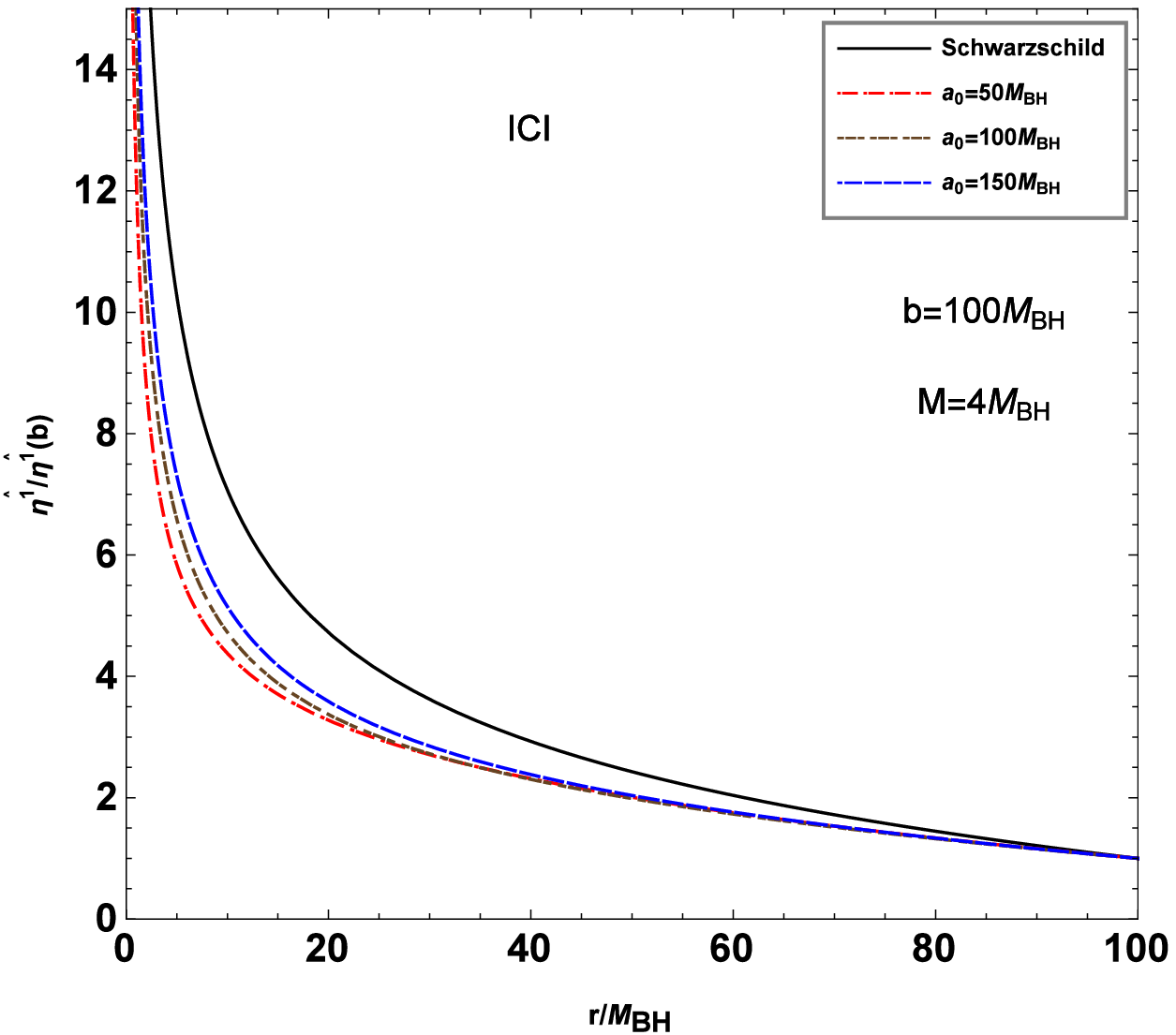}
\includegraphics[width=4.0cm]{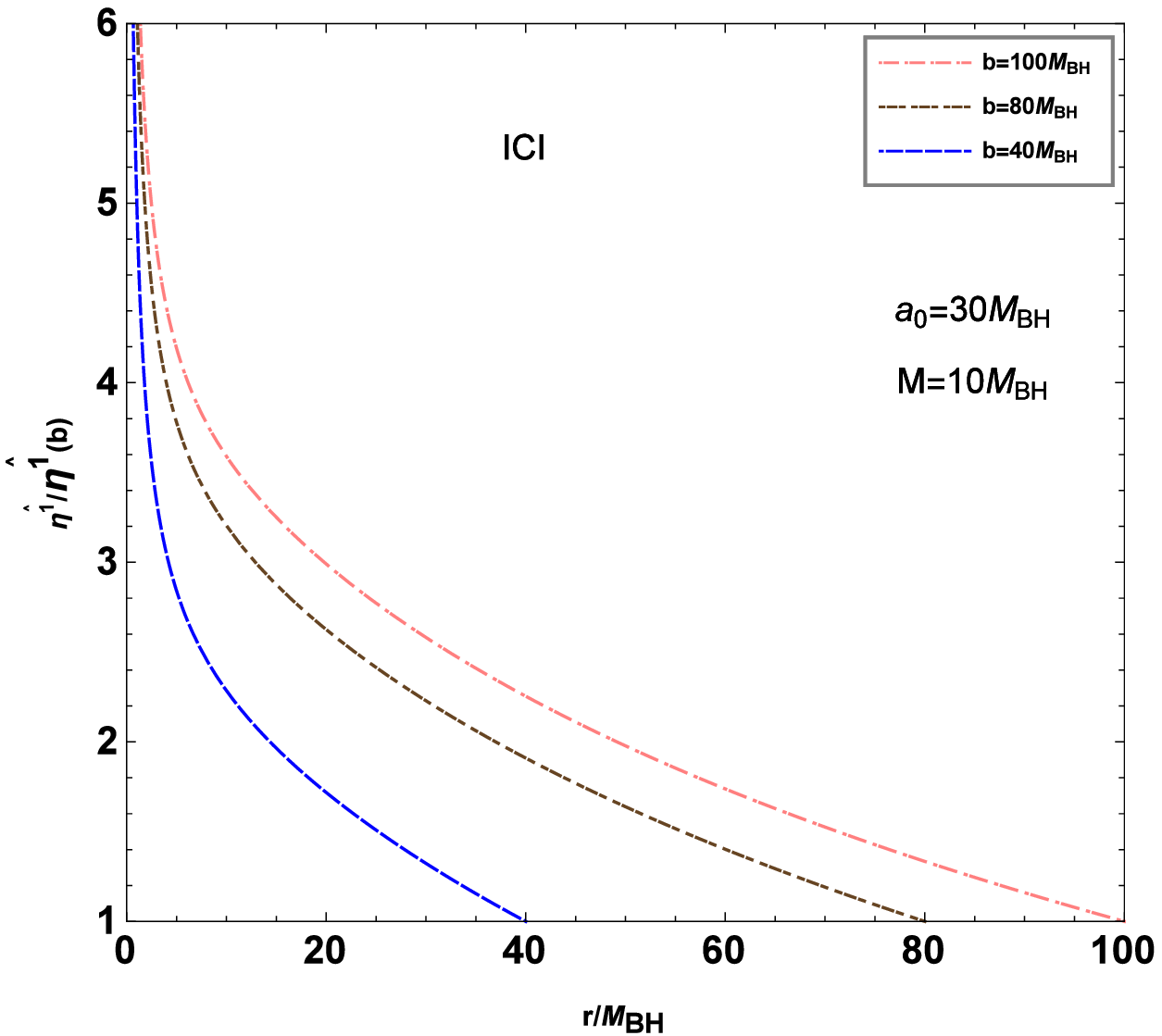}\\
\includegraphics[width=4.0cm]{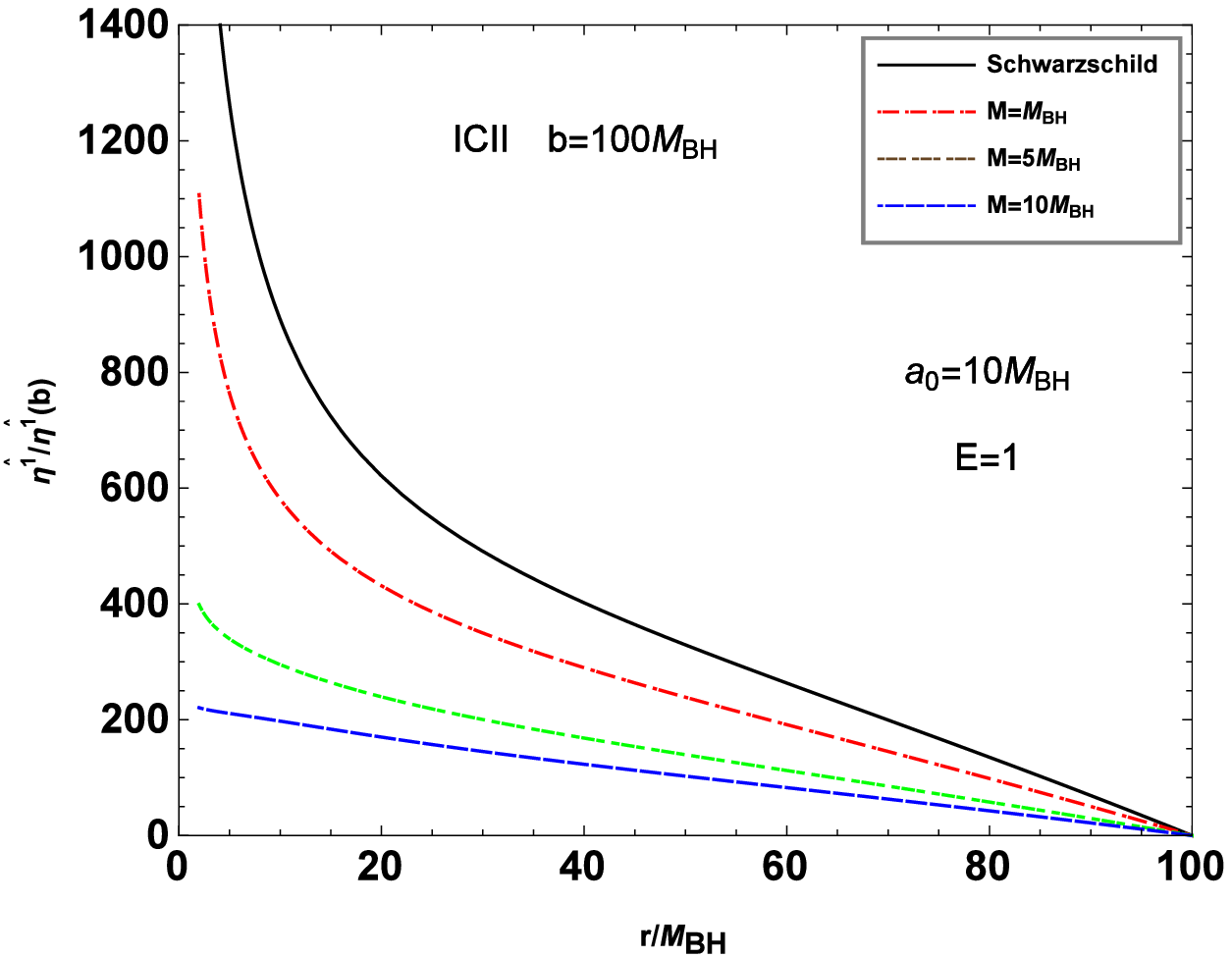}\includegraphics[width=4.0cm]{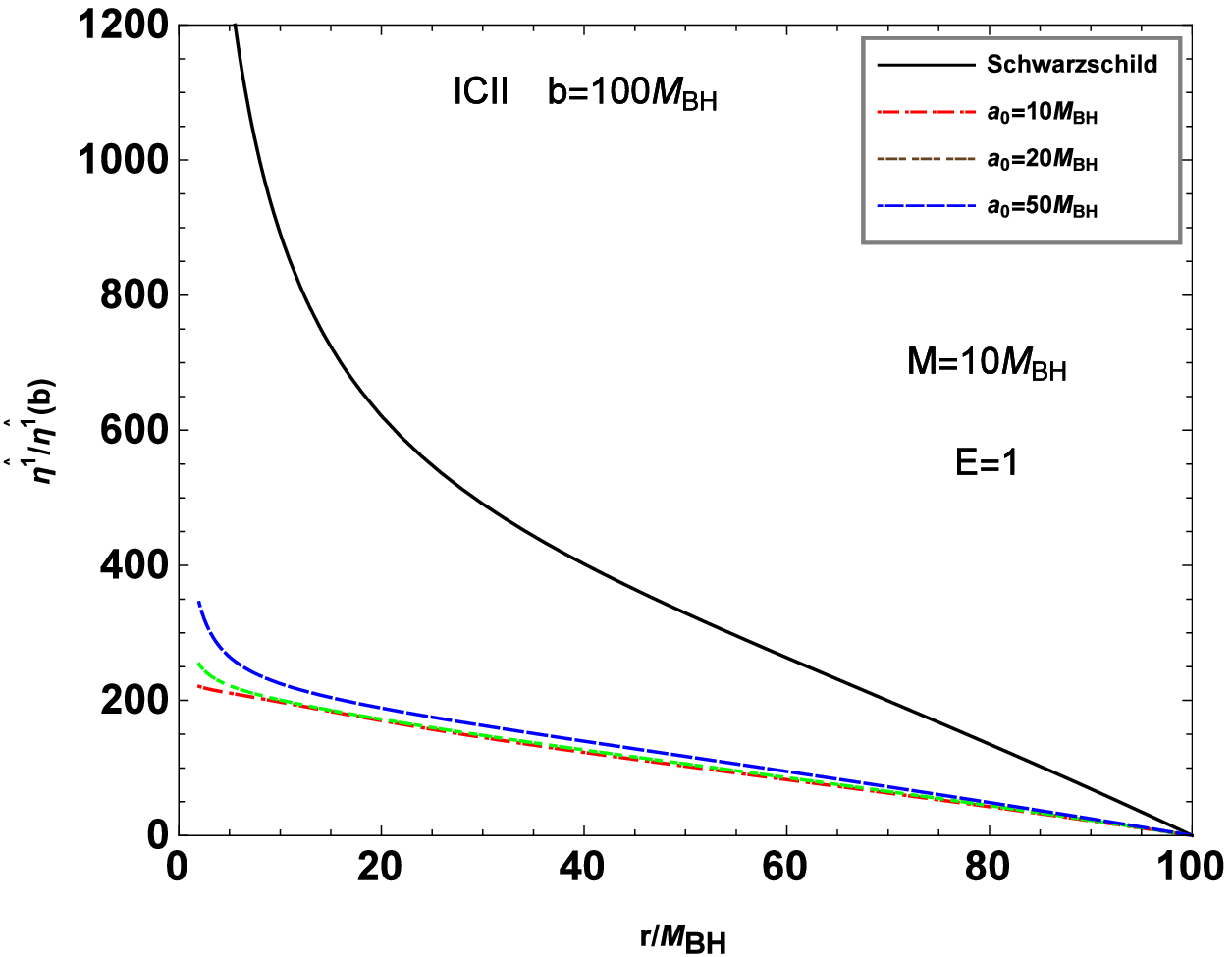}
\includegraphics[width=4.0cm]{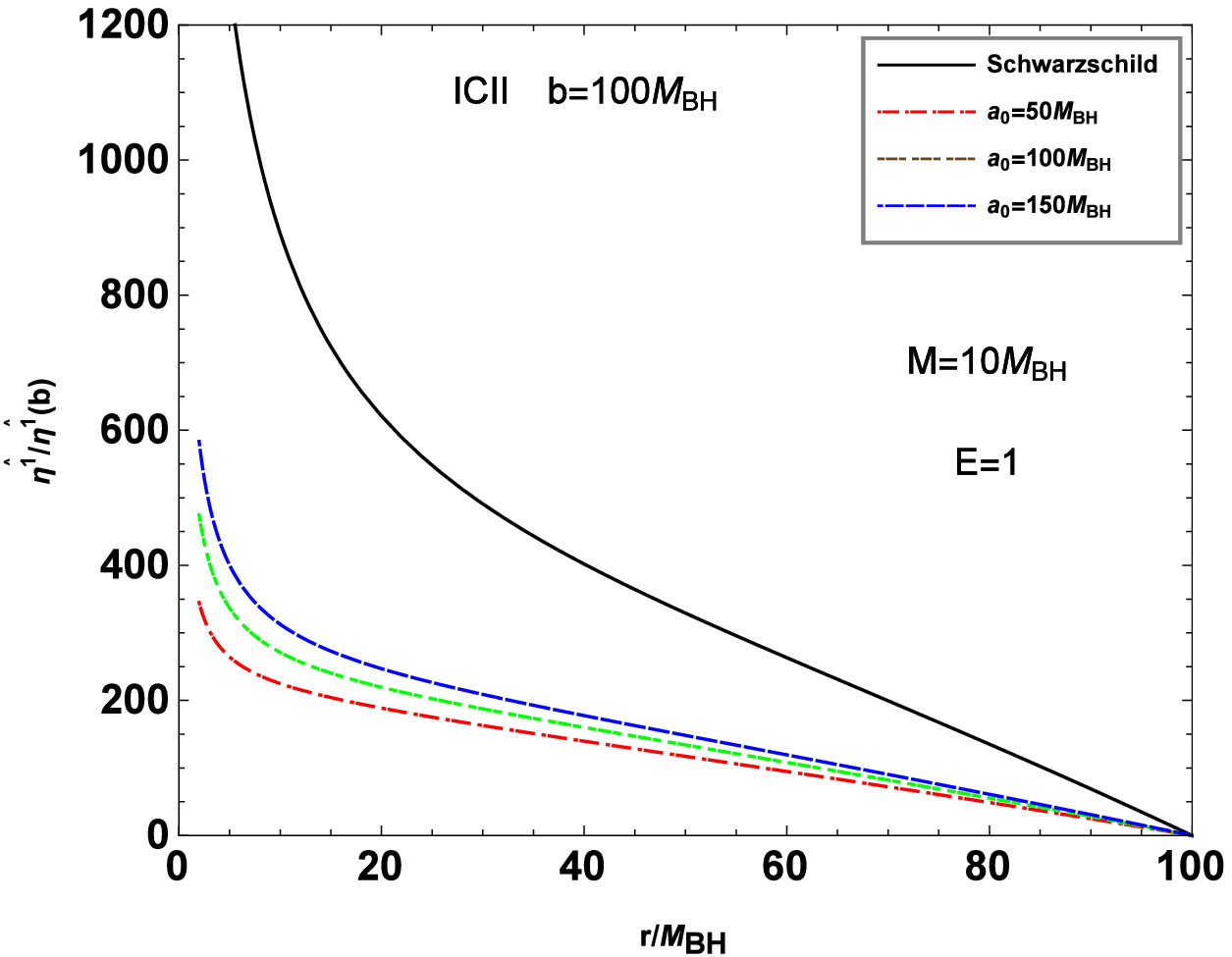}\includegraphics[width=4.0cm]{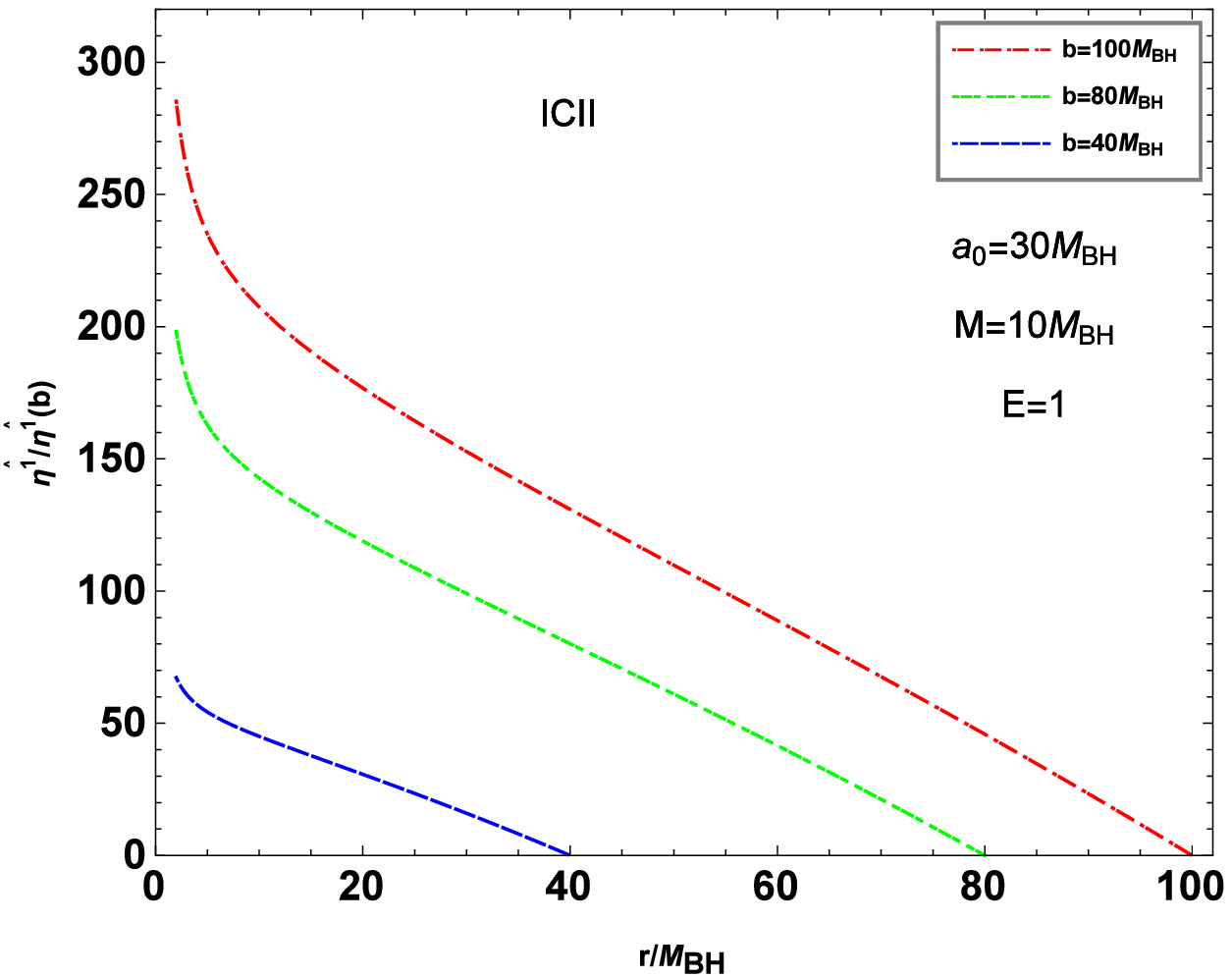}
\\
\includegraphics[width=4.0cm]{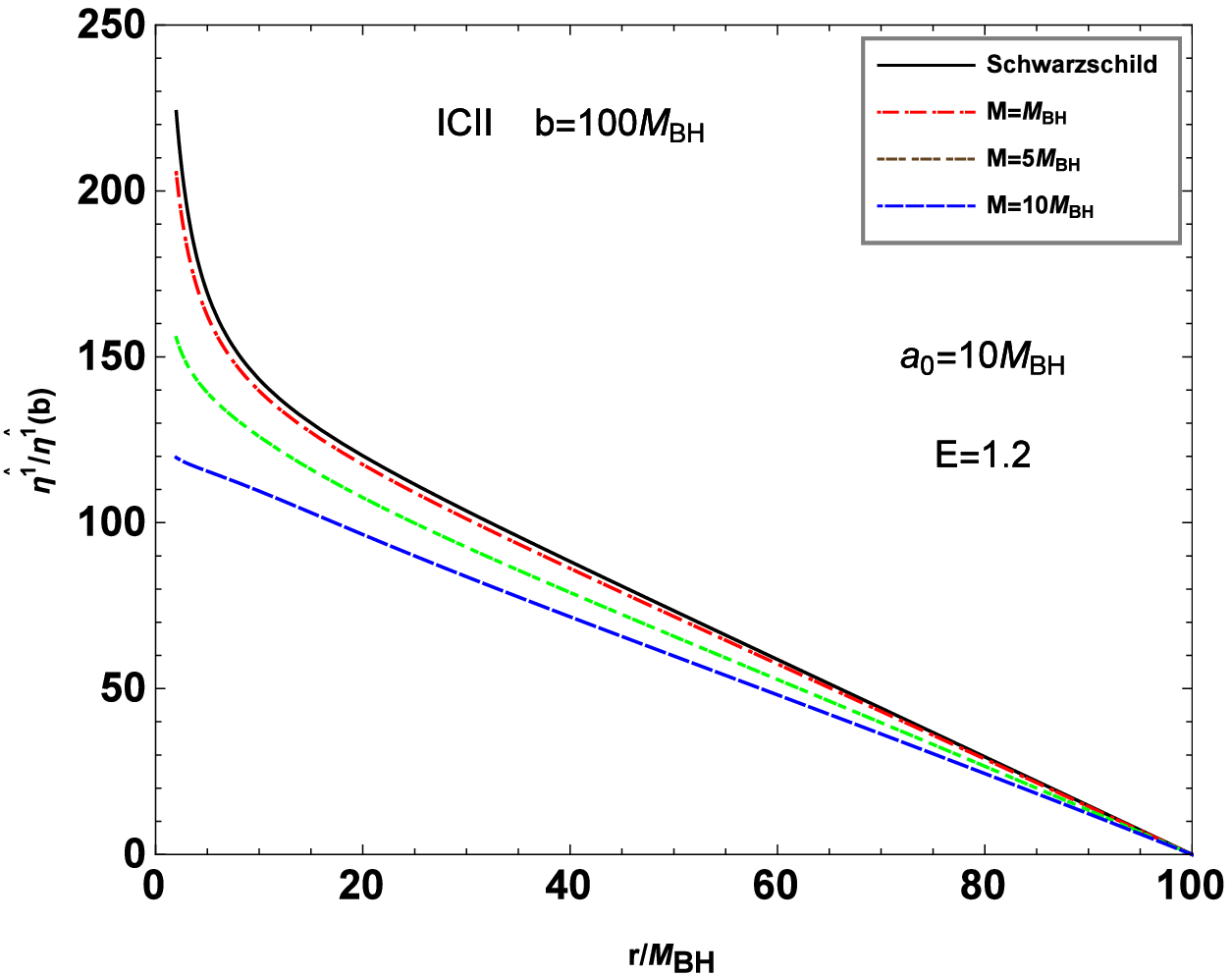}\includegraphics[width=4.0cm]{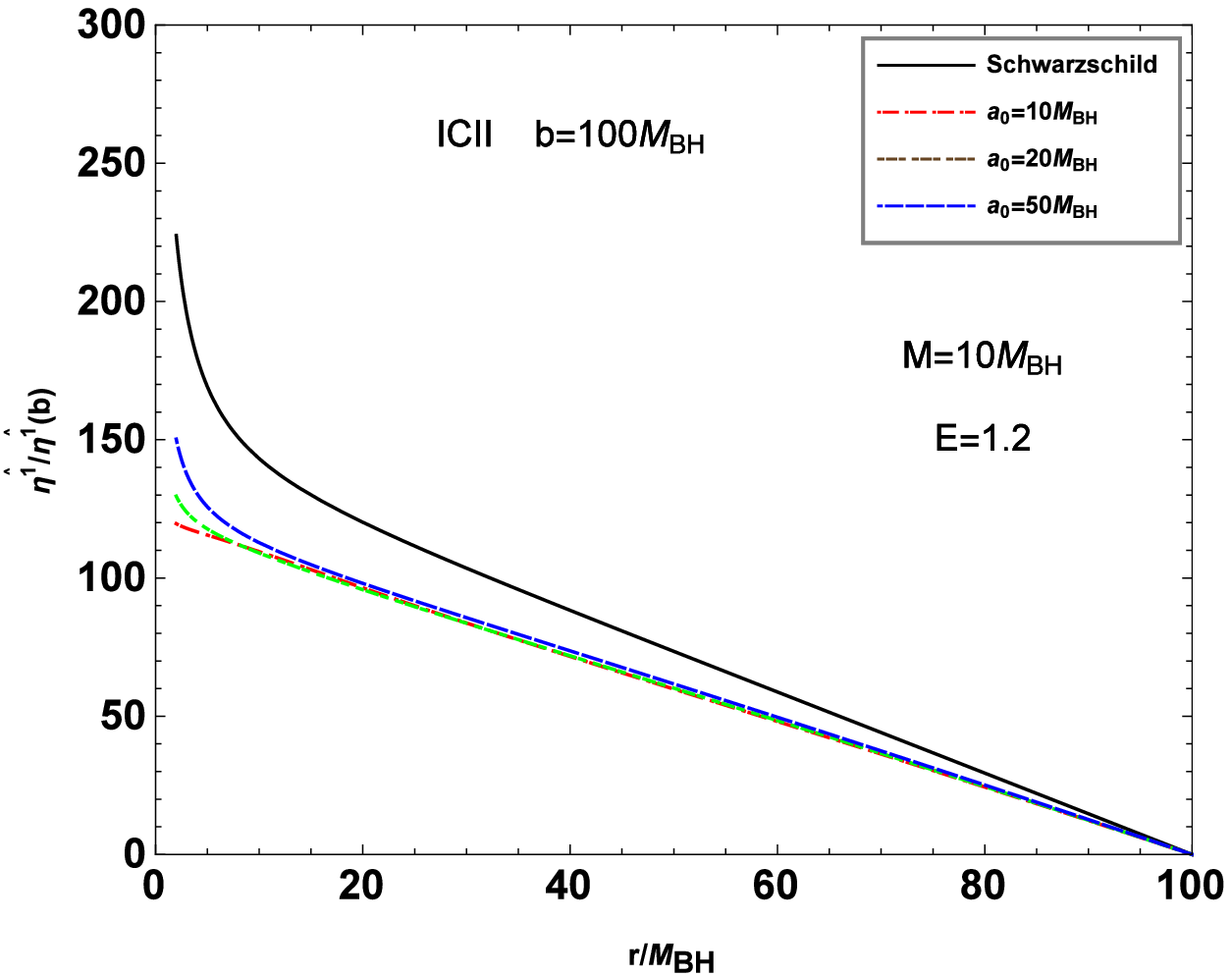}
\includegraphics[width=4.0cm]{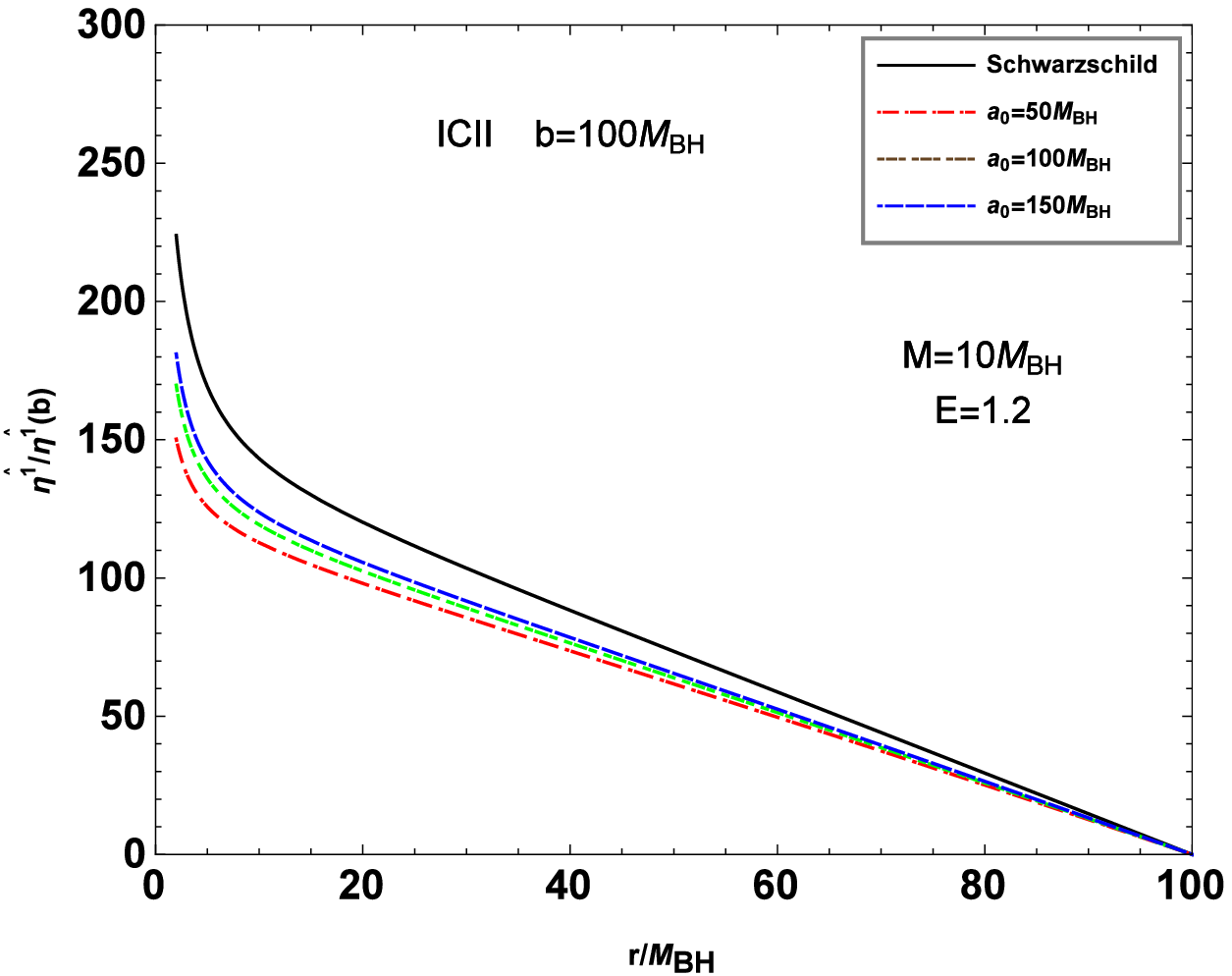}\includegraphics[width=4.0cm]{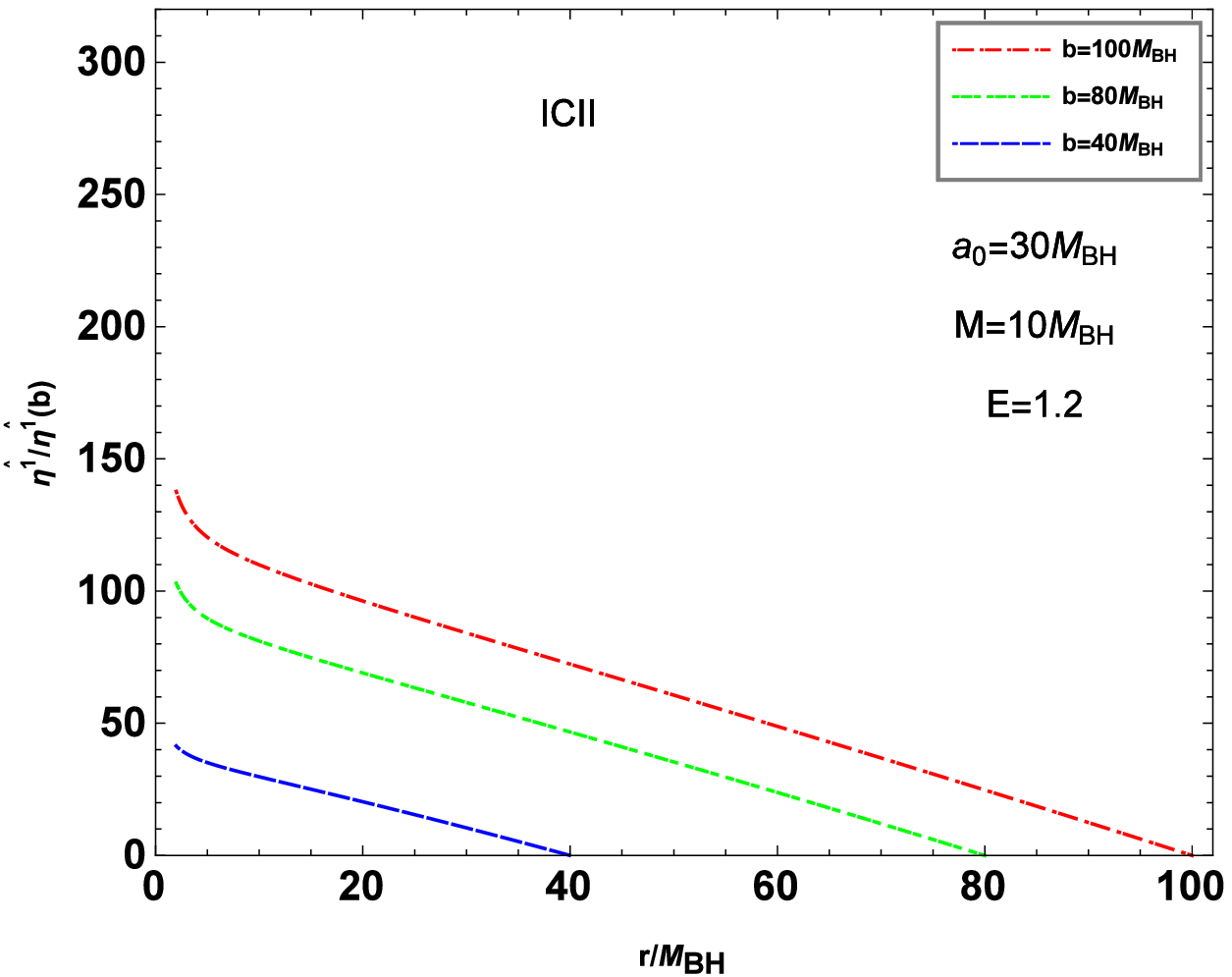}
\caption{ Dependence of the radial components $\eta ^{\hat{1}}$ of the geodesic deviation vector on the dark matter halo mass $M$ and the typical lengthscale $a_0$ in the cases with the initial conditions ICI and ICII. }\label{f5radil}
\end{figure}
Under the initial condition ICII, we find that
\begin{equation}
\eta^{\hat{\mu}'}|_{r=b}=\frac{1}{\dot{r}}|_{r=b}=-\frac{1}{\sqrt{\left(1-\frac{2 m(b)}{b}\right)\left(\frac{E^2}{f(b)}-1\right)}},
\end{equation}
where the energy of the particle-like dust $E$ is not a fixed parameter. The absolute value of $\eta^{\hat{\mu}'}|_{r=b}$ decreases with $E$, which means that the dust energy $E$ will affect the dynamical evolution of the deviation vector in the case with the second kind of initial condition ICII.

In Fig.(\ref{f5radil}), we present change of the radial component of geodesic deviation vector $\eta ^{\hat{1}}$ with the dark matter mass parameter $M$ and  the typical lengthscale $a_0$  in the background of a black hole with dark matter halo (\ref{metric}) with the initial conditions ICI and ICII. As the particle falls from rest at $r=b$, the radial component $\eta ^{\hat{1}}$  increases for different $M$ and $a_0$ under two initial conditions ICI and ICII. The length of $\eta ^{\hat{1}}$and its increasing rate are less than those in the Schwarzschild black hole spacetime. Moreover, the deviation vector component  $\eta ^{\hat{1}}$ monotonously increases with the dark matter parameter $M$. With the increasing of $a_0$, for the first type initial condition ICI, we find that  $\eta ^{\hat{1}}$ increases as the dust lies in the region far from black hole, but first decreases and then increases in the region near the black hole. However, in the case with the second type initial condition ICII, $\eta ^{\hat{1}}$ always increases with $a_0$, which means that the evolution of the deviation vector component $\eta ^{\hat{1}}$ depends on the initial condition. With the increase of $b$, we find that the component  $\eta ^{\hat{1}}$ increases, which is similar to that in the usual static black hole spacetimes. Especially, we find that the particle' energy $E$ also affects the component  $\eta ^{\hat{1}}$, the effects of $M$, $a_0$ and $b$ on the $\eta ^{\hat{1}}$ are suppressed by the particle's energy $E$.
\begin{figure}
\includegraphics[width=5cm]{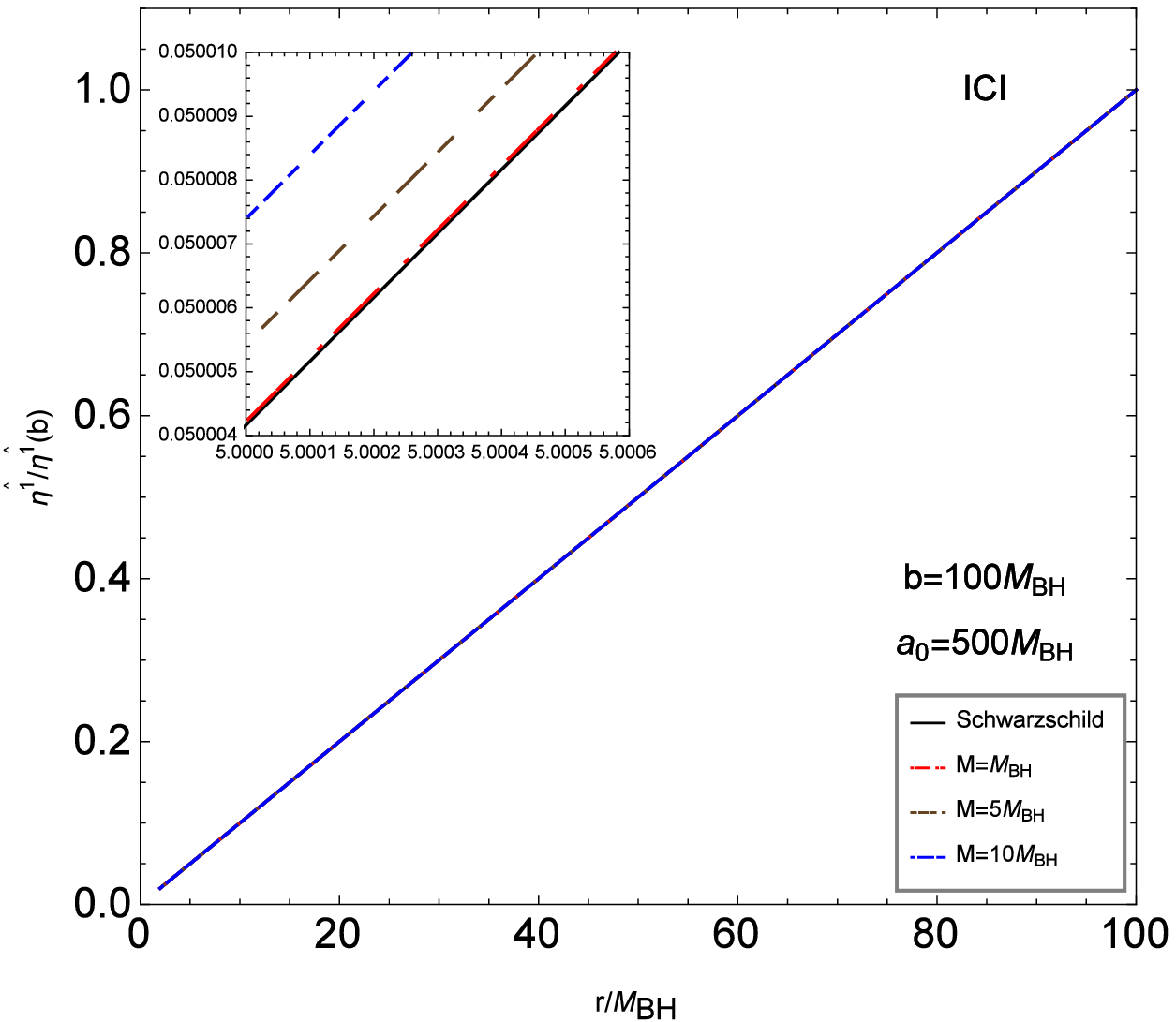}
\includegraphics[width=5cm]{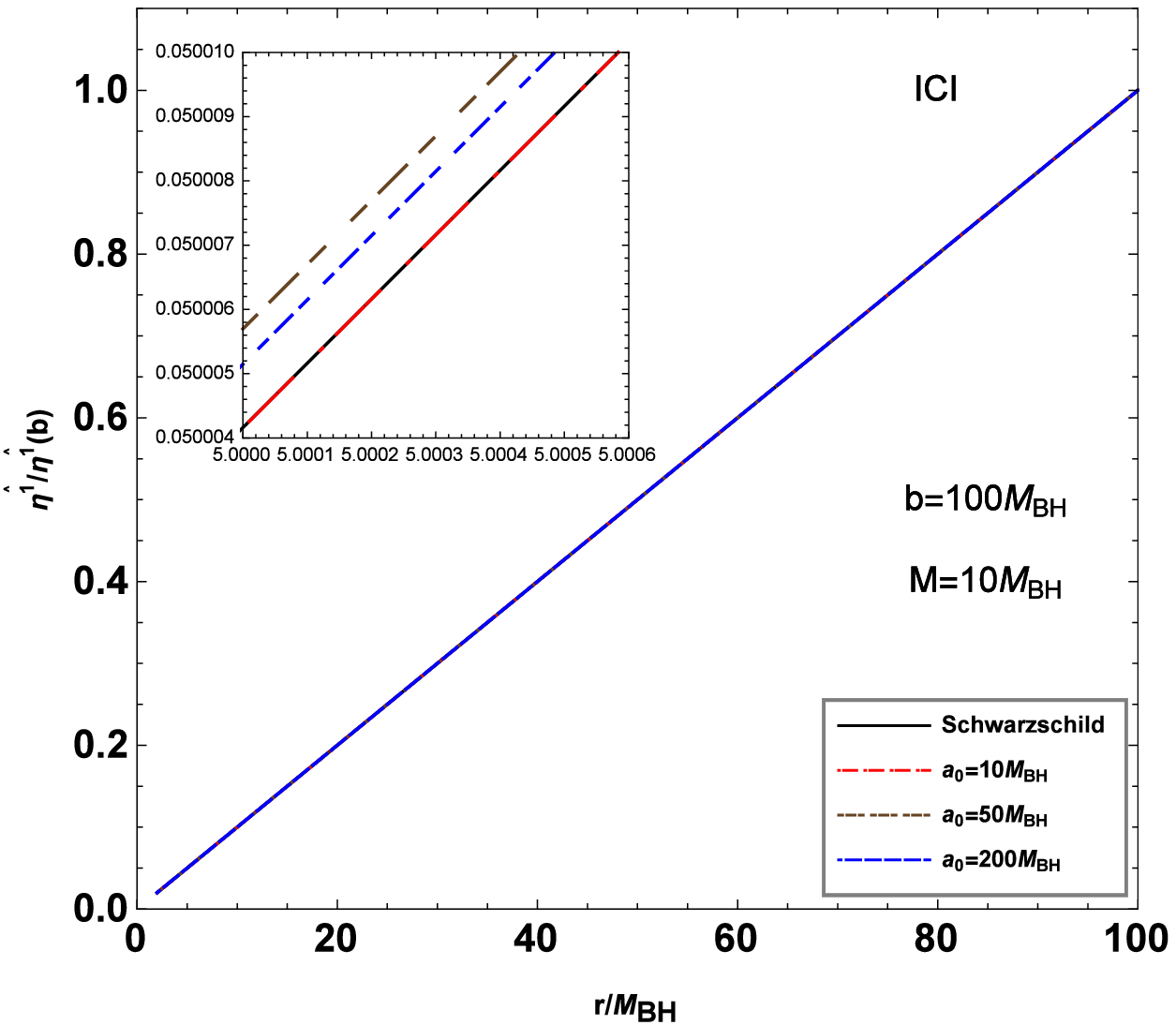}
\includegraphics[width=5cm]{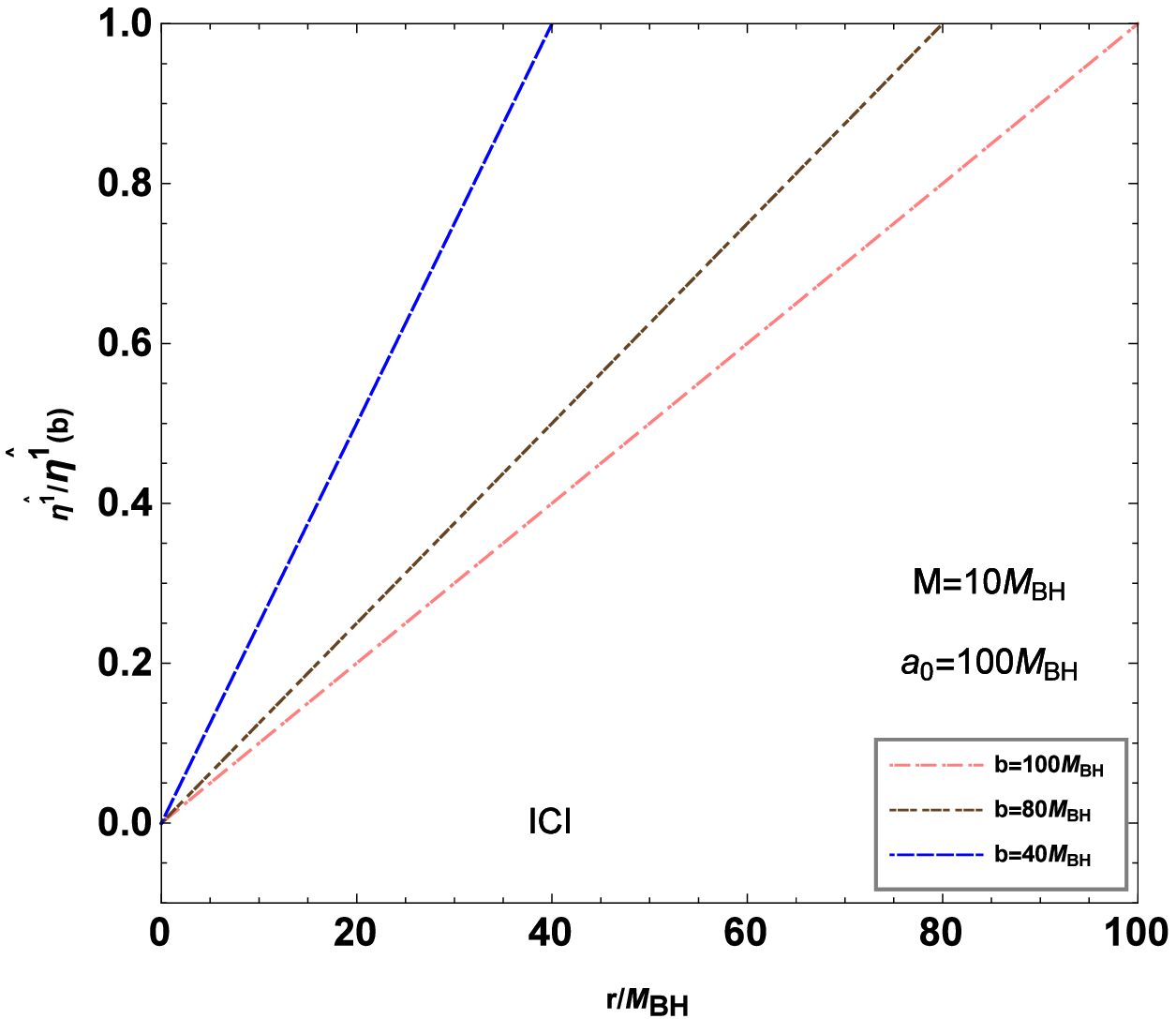}\\
\includegraphics[width=5cm]{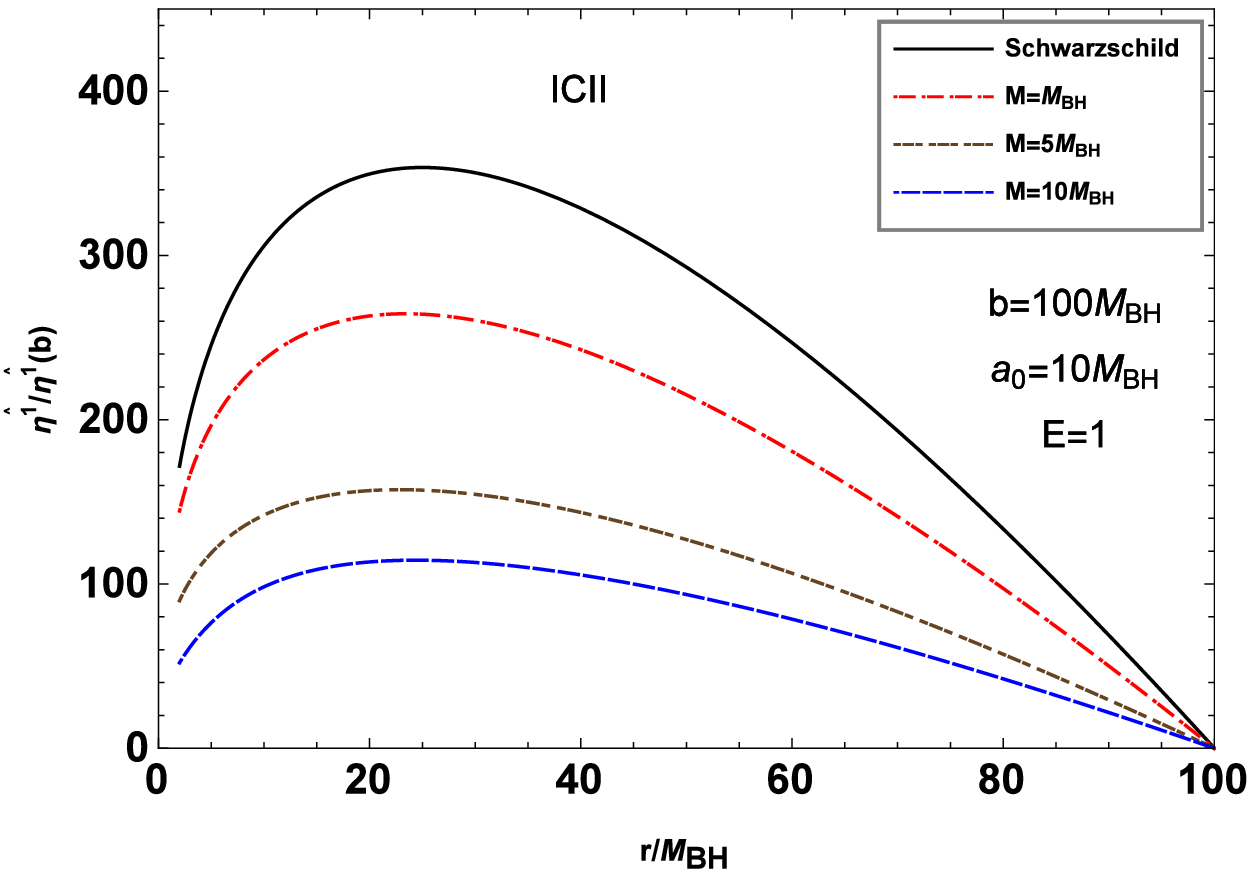}
\includegraphics[width=5cm]{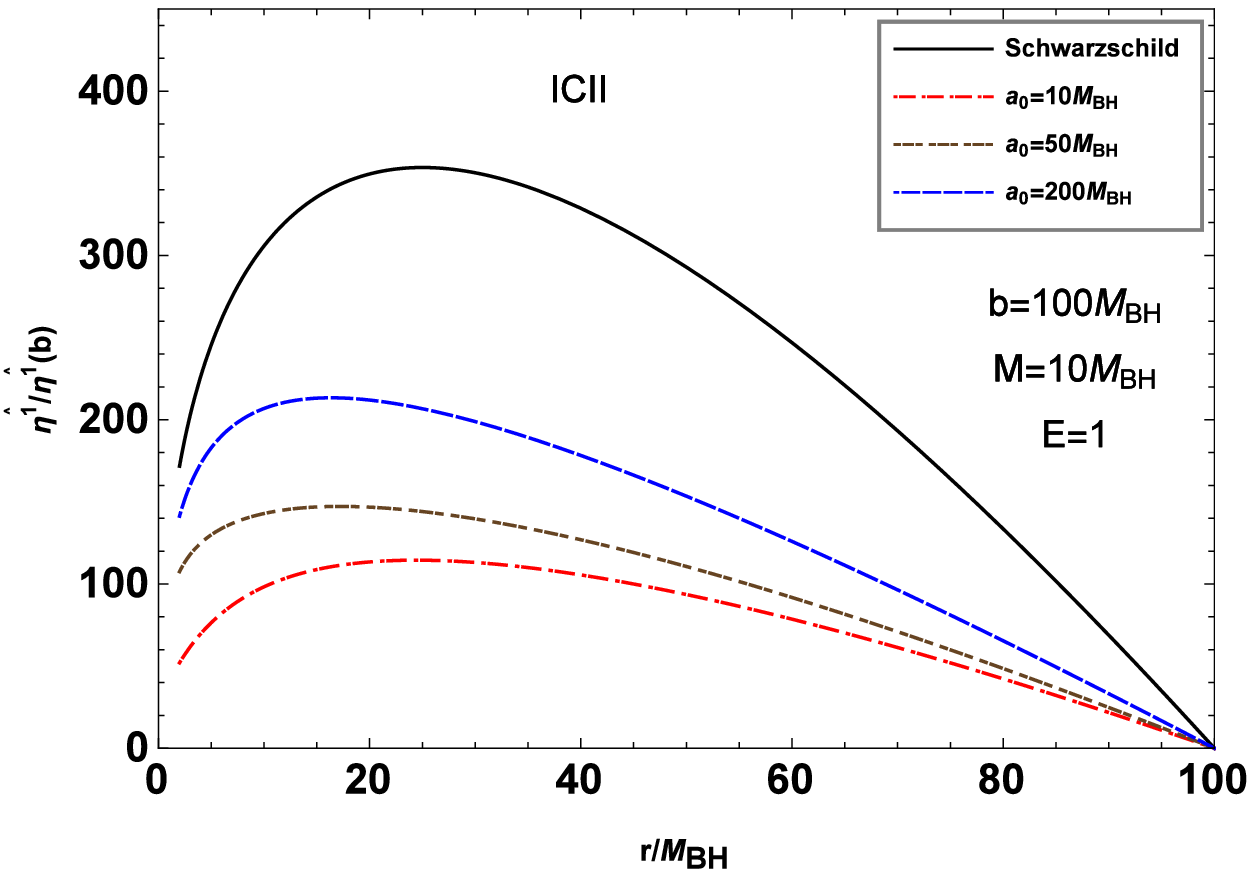}
\includegraphics[width=5cm]{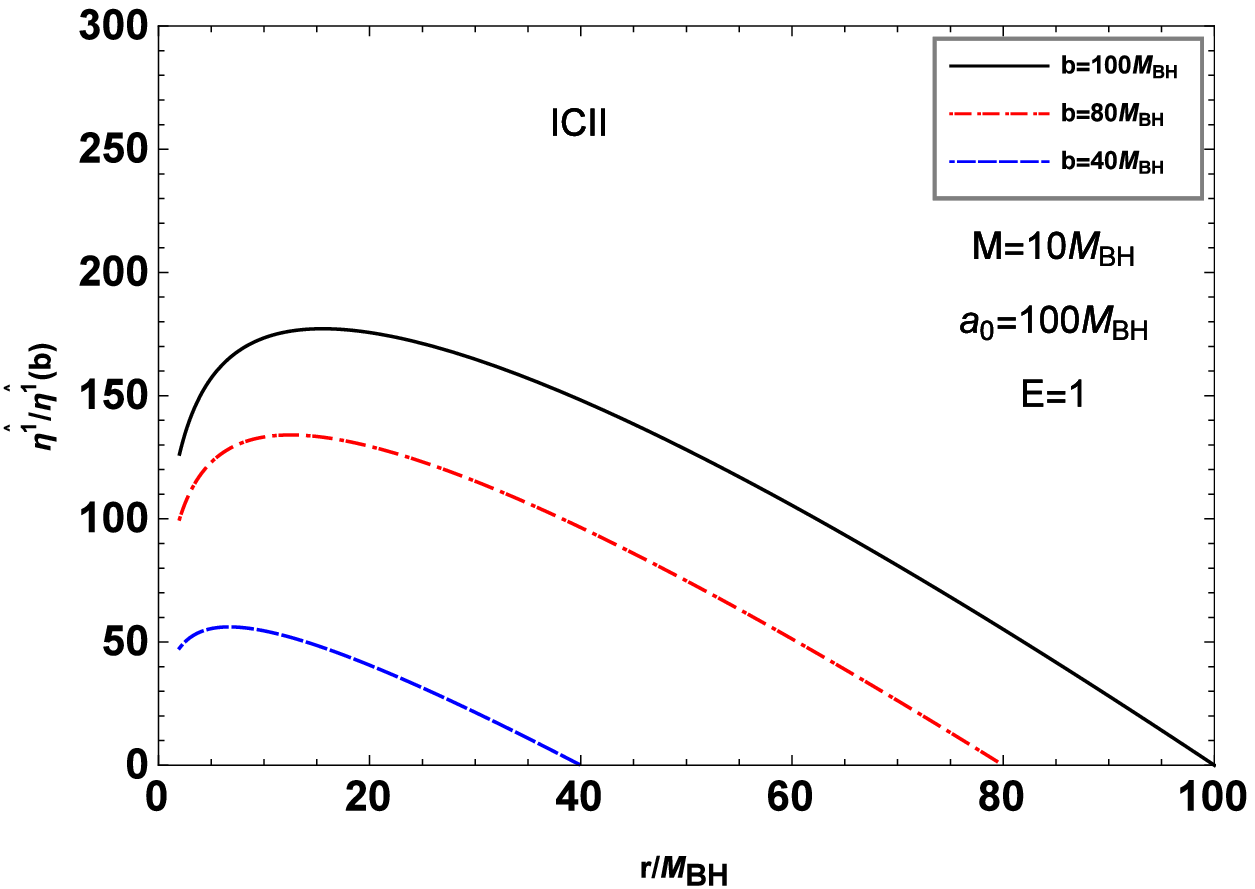}\\
\includegraphics[width=4.8cm]{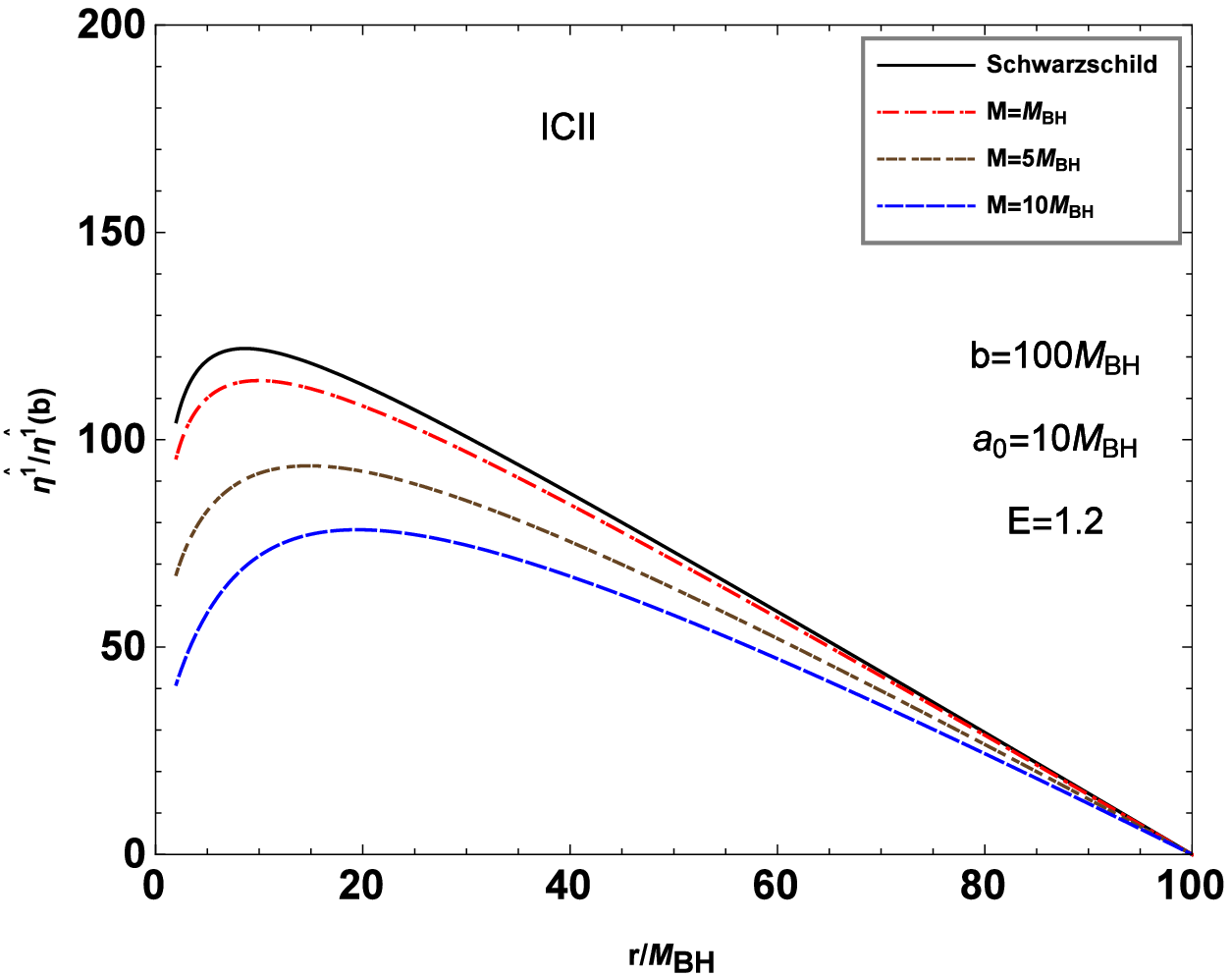}
\includegraphics[width=5.2cm]{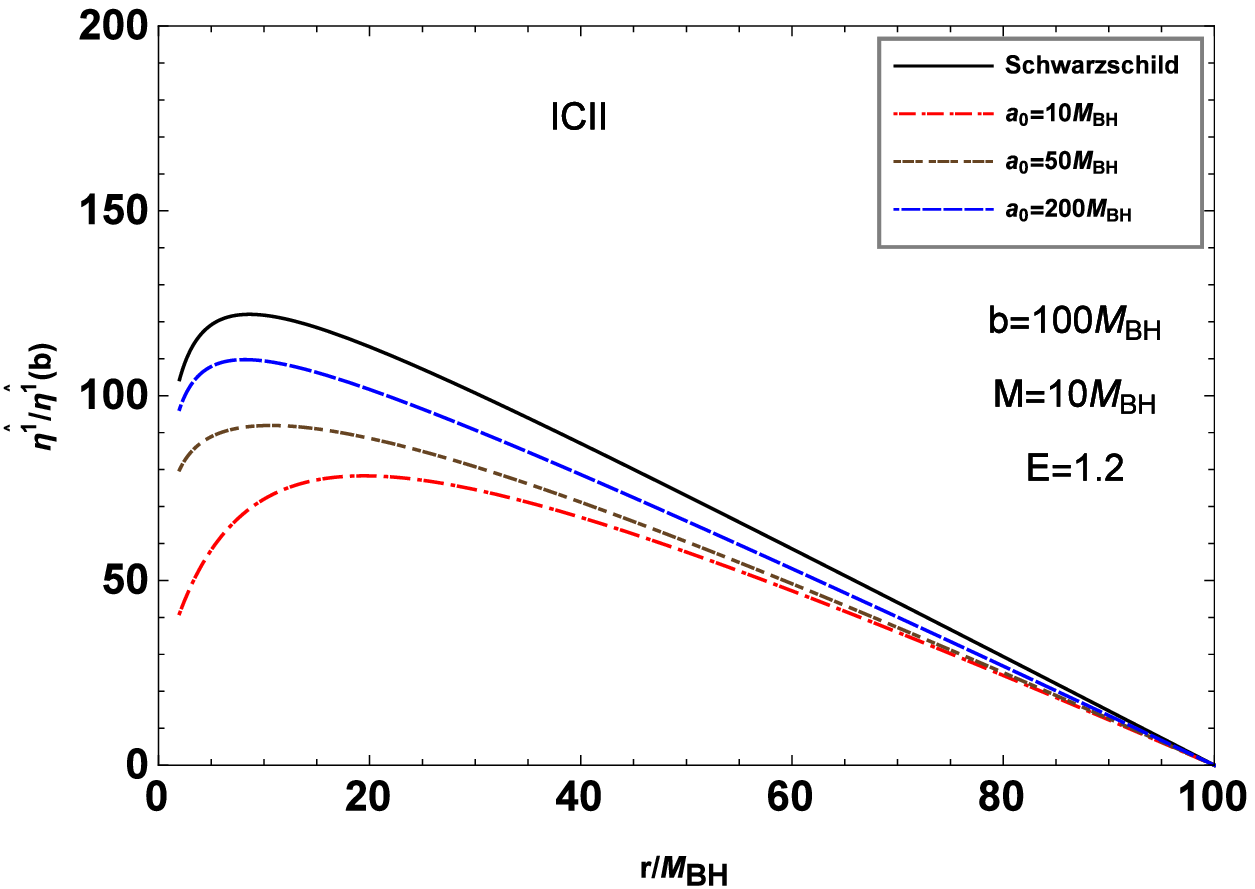}
\includegraphics[width=5.2cm]{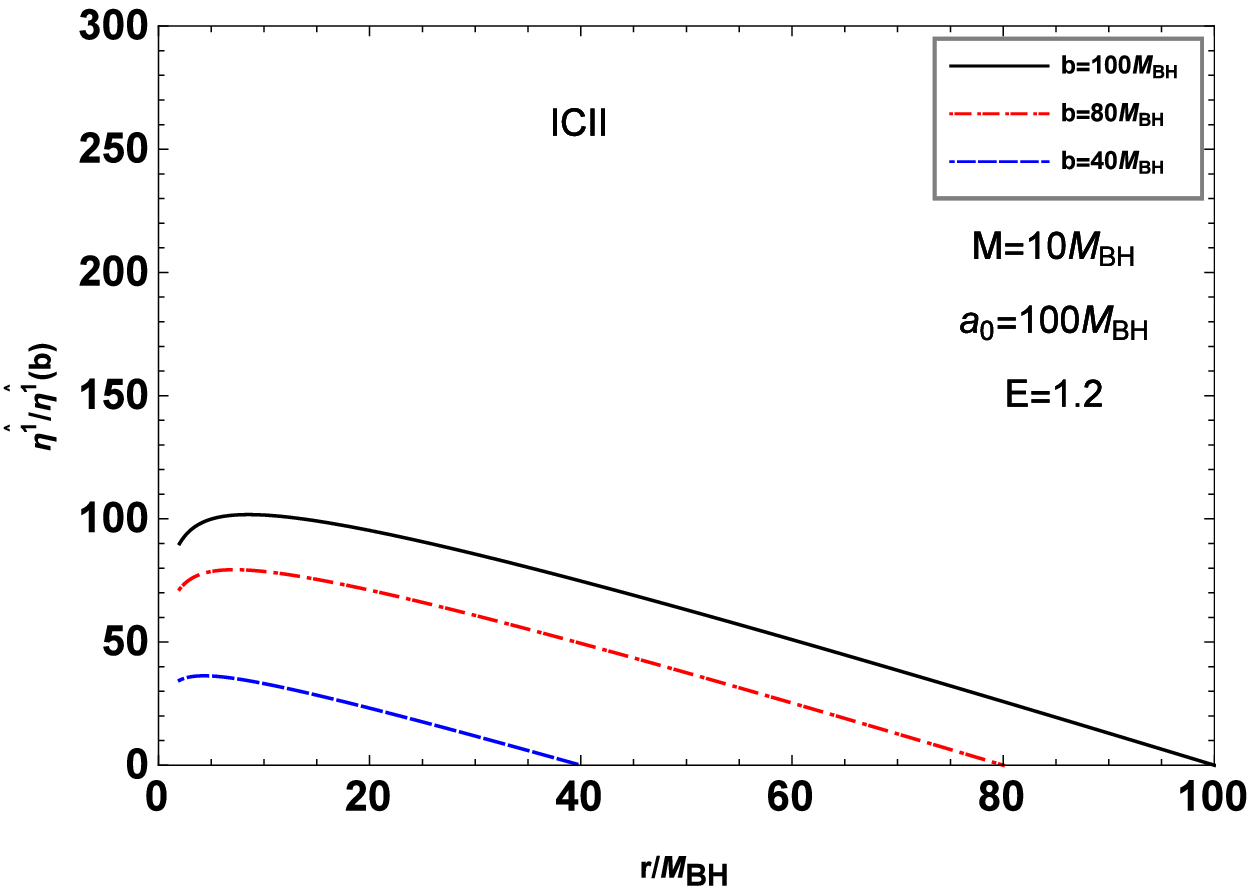}
\caption{ Dependence of the angular components $\eta ^{\hat{i}}$  of the geodesic deviation vector on the dark matter halo mass $M$ and the typical lengthscale $a_0$ in the cases with the initial conditions ICI and ICII. }\label{f6angu}
\end{figure}

Fig.(\ref{f6angu}) shows the dependence of the angular components $\eta ^{\hat{i}}$  of the geodesic deviation vector on the dark matter halo mass $M$ and the typical lengthscale $a_0$.
With in the initial condition ICI, $\eta ^{\hat{i}}$ increases with $M$, but first increases and then decreases with $a_0$. However, the effects of $M$ and $a_0$ on the components $\eta ^{\hat{i}}$ is very tiny in this case. Moreover, for the fixed $M$ and $a_0$, the angular components $\eta ^{\hat{i}}$ decreases with $b$.
For the initial condition ICII, as the body falls from $r=b$, one can find that there is a peak in the change curve of the component $\eta ^{\hat{i}}$ with $r$,  and the peak value of $\eta ^{\hat{i}}$ and its corresponding position depend on the parameters $M$, $a_0$ and $b$. Moreover, the angular components $\eta ^{\hat{i}}$ is a decreasing function of $M$, but an increasing function of $a_0$ and $b$ in this case, which means that the dynamical behaviors of the particle-like dust differs from those in the case with initial condition ICI.  Like the radial component $\eta ^{\hat{1}}$, the effects of $M$, $a_0$ and $b$ on the $\eta ^{\hat{i}}$ are also suppressed by the particle's energy $E$ under the initial condition ICII. Thus, the tidal effects of dark matter halo depend on the dark matter mass $M$ and  the typical lengthscale of the galaxy $a_0$, but also on the particle-like dust's initial condition.

\section{Summary}
We have investigated the tidal forces and geodesic deviation motion in the spacetime of a black hole in the galaxy with dark matter halo (\ref{metric}). Our results show that the tidal force and geodesic deviation
motion depend on the dark matter halo mass $M$ and the typical lengthscale $a_0$ of galaxy. The effect of $a_0$ on  tidal force is opposite to that of $M$ in the background of a black hole with dark matter halo. The main reason is that the dark matter density around black hole increases with the mass parameter $M$, but decreases with the scale parameter $a_0$.  For the radial tidal force, with the increasing mass $M$ of dark matter, it increases in the region far from the black hole, but decreases in the region near black hole. The main reason is that for the test particle in the far region the gravity arising from the dark matter around black hole has the same direction as that of black hole at the center of galaxy, but they are in the opposite directions for the particle in the near region.  The angular tidal force is negative for all $M$ and $a_0$,  its absolute value of angular tidal force monotonously increases with the dark matter halo mass. Especially, we also find that the angular tidal force also depends on the particle's energy $E$ and  the effects of $M$ and $a_0$ become more distinct for the test particle with high energy $E$, which is different from those in the usual static black hole spacetimes where the angular tidal force is independent of the test particle's energy $E$.

We also present the change of geodesic deviation vector in the spacetime of a black hole in the galaxy with dark matter halo under two conditions ICI and ICII. The dependence of geodesic deviation vector on the initial position parameter $b$ is similar to that in other black hole spacetimes. The radial deviation vector component  $\eta ^{\hat{1}}$ monotonously increases with the dark matter parameter $M$. With the increasing of $a_0$, for the first type initial condition ICI, $\eta ^{\hat{1}}$ increases as the dust lies in the region far from black hole, but first decreases and then increases in the region near the black hole. However, in the case with the second type initial condition ICII, $\eta ^{\hat{1}}$ always increases with $a_0$.
For the angular components $\eta ^{\hat{i}}$, with the initial condition ICI, it increases with the dark matter mass parameter $M$, but first increases and then decreases with $a_0$. However, the effects of $M$ and $a_0$ on the components $\eta ^{\hat{i}}$ is very tiny in this case. For the initial condition ICII,  the angular components $\eta ^{\hat{i}}$ is a decreasing function of $M$, but an increasing function of $a_0$ and $b$ in this case, which means that the dynamical behaviors of the particle-like dust differs from those in the case with initial condition ICI.
Especially, we find that the particle' energy $E$ also affects the dynamical evolution of geodesic deviation vector in the black hole spacetime, and the effects of $M$, $a_0$ and $b$ on the  deviation vector are suppressed by the particle's energy $E$. These behaviors of tidal forces and geodesic deviation vector could help us to
understand tidal effects and dark matter halo around a galactic black hole.

\texttt{}
\section{\bf Acknowledgments}

This work was  supported by the National Natural Science
Foundation of China under Grant No.11875026, 11875025, 12035005, and 2020YFC2201400.

\bibliography{tidal-dark-matter}

\end{document}